\PassOptionsToPackage{quiet}{fontspec}
\documentclass[a4paper,12pt]{article}
\usepackage[a4paper,top=2.54cm, bottom=2.54cm, left=3.2cm, right=3.2cm]{geometry}
\usepackage{amsmath,amssymb,amsthm,bm}
\usepackage{makecell}
\usepackage{graphicx}
\usepackage{float}
\usepackage{color}

\usepackage{natbib}
\usepackage[sort&compress]{gbt7714}
\usepackage{indentfirst}
\usepackage{multirow}
\usepackage{threeparttable}

\newtheorem{theorem}{Theorem}

\newtheorem{lemma}{Lemma}

\usepackage{authblk}
\usepackage{algorithm}
\usepackage{algpseudocode}
\usepackage{booktabs}
\floatname{algorithm}{Algorithm}
\title{\textbf{A Local Modal Outer-Product-Gradient Estimator for Dimension Reduction}}

\author[1]{Zheng Li \thanks{First author: liz768@nenu.edu.cn}}

\author[1]{Chong Ding}

\author[1]{Wei Gao \thanks{Corresponding author: gaow@nenu.edu.cn}}

\affil[1]{Key Laboratory for Applied Statistics of MOE, School of Mathematics and Statistics, Northeast Normal University, Changchun 130024, China.}

\begin{document}

\date{}

\maketitle{}

\maketitle

\begin{abstract}
Sufficient dimension reduction (SDR) is a valuable approach for handling high-dimensional data. 
Outer Product Gradient (OPG) is an popular approach. However, because of focusing the mean regression function, OPG may ignore some directions of central subspace (CS) when the distribution of errors is symmetric about zero. The mode of a distribution can provide an important summary of data. 
A Local Modal OPG (LMOPG) and its algorithm through mode regression are proposed to estimate the basis of CS with skew errors distribution. The estimator shows the consistent and asymptotic normal distribution under some mild conditions.
Monte Carlo simulation is used to evaluate the performance and demonstrate the efficiency and robustness of the proposed method.

\textbf{Keywords:} sufficient dimension reduction; central subspace; outer product gradient; mode regression.
\end{abstract}

\section{Introduction}
Consider $\bm{X}$ is a $p$-dimensional covariate vector, $\textsl{y}$ is a  univariate response variable. \cite{Li1991} proposed a model that the regression relation between $\textsl{y}$ and $\bm{X}$ is unknown, and all the information between  $\bm{X}$ and $\textsl{y}$ can be carried from $d$-dimensional projections $\left(\beta_{01}^\top \bm{X},\beta_{02}^\top \bm{X},\cdots,\beta_{0d}^\top \bm{X} \right)$, the $\beta_{i}$'s are unknown column vectors in $\mathbb{R}^{p}$ and denoted by $\text{B}_0^\top \bm{X}$. This can be expressed as
\begin{equation}
\label{M1}
\textsl{y}=g\left(\beta_{01}^\top \bm{X},\beta_{02}^\top \bm{X},\cdots,\beta_{0d}^\top \bm{X},\varepsilon\right),
\end{equation}
where $g$ is a unknown link function,  and $\varepsilon$ is independent of $\bm{X}$. This is equivalent to 
\begin{equation}
\label{y||X}
    \textsl{y}\perp \!\!\! \perp \bm{X} \mid \text{B}_0^\top \bm{X},
\end{equation}
that is, $\textsl{y}$ depends on $\bm{X}$ only through $\text{B}_0^\top \bm{X}$. The basic problem of sufficient dimension reduction (SDR) is to find the basis matrix $\text{B}_0$, that is the column space of matrix $\text{B}$ and denoted by $\textup{Span}(\text{B}_0)$. The $\textup{Span}(\text{B}_0)$ is often known as the effective dimension reduction (EDR) space. Notice that $\text{B}_0$ always exists because $\text{B}_0$ degenerates into an identity matrix when $k=p$, and it is not unique because $\textsl{y}\perp \!\!\! \perp \bm{X} \mid \text{P}\text{B}_0^\top \bm{X}$ for any nonsingular matrix $\text{P}$. Due to the uniqueness of $\text{B}_0$, \cite{Cook1998} established the central subspace (CS), and written as $\bm{S}_{\textsl{y}\mid \bm{X}}$, which is the intersection of all subspace $\textup{Span}(\text{B})$. $\bm{S}_{\textsl{y}\mid \bm{X}}$ is the smallest dimension reduction space and is unique. When our primary
interest is the mean function of regression, the objective of SDR is to find the basis matrix $\text{B}_0$ such that
\begin{equation}
\label{y||E(y|X)}
    \textsl{y}\perp \!\!\! \perp E(\textsl{y}\mid \bm{X}) \mid \text{B}_0^\top \bm{X}.
\end{equation}
The basis matrix satisfying \eqref{y||E(y|X)} is called mean dimension reduction subspace. Similarly, the central mean subspace (CMS) is established by \citep{cook2002dimension} and is denoted by $\bm{S}_{E(\textsl{y}\mid \bm{X})}$.
It is obviously that $\bm{S}_{E(\textsl{y}\mid \bm{X})}\subseteq \bm{S}_{\textsl{y}\mid \bm{X}}$. Because \eqref{y||X} is concerned about condition distribution $\textsl{y}\mid\bm{X}$, \eqref{y||E(y|X)} is concerned about the mean of condition distribution.

The exiting methods of SDR can be divided into two types: inverse regression and forward regression. Just as its name implies, the inverse regression is the condition distribution of giving the response variable $\textsl{y}$, i.e. $\bm{X}\mid \textsl{y}$. The well known inverse method called Sliced Inverse Regression (SIR) is first proposed by \citep{Li1991}. Under the assumption of elliptical distribution, the curve of inverse regression $E(\bm{X}\mid\textsl{y} )$ falls in $\bm{S}_{\textsl{y}\mid \bm{X}}$. So, the eigenvector of $\text{Cov}(E(\bm{X}\mid\textsl{y} ))$ can be used as the basis of $\bm{S}_{\textsl{y}\mid \bm{X}}$. In addition, there are other inverse regression methods, such as: Sliced Average Variance Estimate (SAVE) \citep{cook1991SAVE}, Parametric Inverse Regression  \citep{PIR2001}, Canonical Correlation Estimator  \citep{fung2002}, Contour Regression (CR) \citep{libing2005}, Inverse Regression Estimator (IRE) \citep{cook2005}, principal fitted components \citep{cook2007}, Directional Reduction \citep{li2007directional}, Elliptically Contour Inverse Rredictors \citep{bura2015sufficient},  Generalized
Kernel-based Inverse Regression \citep{xie2020generalizedkernel} and Elliptical sliced inverse regression (ESIR) \citep{chen2022high}.

Another forward regression methods of SDR  focus on the condition distribution by giving $\bm{X}$, that is, $\textsl{y}\mid \bm{X}$. \cite{li1989regression} indicate that Ordinary Least Squares (OLS) can be considered as a dimension reduction method. Principle Hessian Directions (PHD) are proposed by \citep{li1992principal} concentrate on the second-order partial derivative of regression mean function, i.e., the Hessian matrix of $E(\textsl{y}\mid \bm{X})$, and transform it to an easily solvable form by Stein's lemma. Besides, some works related to PHD are Iterative Hessian Transformation (IHT) \citep{cook2002dimension} and further developed in \citep{cook2004determining}, Generalized PHD (GPHD)  \citep{chen2018generalizedprinciple} and Adjusted PHD (APHD) \citep{luo2018secondorder}. These extensions are mainly done by relaxing the assumptions about the distribution of $\bm{X}$. There are another SDR approaches that have less restrictive distribution of predictor variables $\bm{X}$. Such as Minimum Average Variance Estimator (MAVE) \citep{xia2002adaptive}, Ensemble of Minimum Average Variance Estimator (EMVAE)\citep{yin2011sufficient}, roubust sparse MAVE for variable selection \citep{yao2013robust}, Semiparametric Dimension Reduction Methods (SDRM) \citep{ma2012semiparametric}, Outer-Product-Gradient Method (OPG) \citep{xia2002adaptive}. 

OPG is a popular method for estimating CMS, which focus on the first-order partial derivative of $E(\textsl{y}\mid \bm{X})$ and combines the local linear regression \citep{fan2018local}. Furthermore, \cite{opg2007}, \cite{2014Xiaadaptive} and \cite{kang2022forward} obtain the whole estimation of CS based on OPG. Since OPG uses the least squares criterion, OPG may be not robust under heavy tailed error distributions. The mode of a distribution provides an important summary of data. 
\cite{yao2012local} proposed the
local modal regression (LMR), which models the
mode instead of the mean function, it has better performance when heavy tailed error distributions and further developed in nonparametric modal regression (LPMR) \citep{xiang2022nonparametric}. \cite{rekabdarkolaee2017robust} has emphasized this point above and proposed Local Mode MAVE (LMMAVE). There are many literature have focused on idenfing modes of pupulation distributions for low-dimensional data, including LMR and LPMR.
 \cite{yao2014new} propose a new regression which is called modal linear regression (MODLR) to explore high-dimensional data. MODLR models the conditional mode of $\textsl{y}$ given $\bm{X}$ as a linear
function of $\bm{X}$. An expectation–maximization algorithm are proposed to estimate the regression coefficients and its asymptotic properties with the skew error density is well.

In this paper, we combine the idea of OPG, LMR and MODLR to introduce a new approach called Local Modal outer product gradient (LMOPG) for SDR.  LMOPG models the conditional mode of a response $\textsl{y}$ given a set of predictors $\bm{X}$ as a nonlinear function of $\bm{X}$, which has a low-dimensional linear combination structure, i.e., $\textup{B}^\top \bm{X}$. By performing a Taylor expansion of this nonlinear function, the first-order partial derivative is the linear combinations of columns of matrix $\textup{B}$, and then we perform spectral decomposition for the second moment of the first-order partial derivative. We propose LMRPG algorithm in order to estimate the basis of CS. LMOPG can take full advantage of the mode of distribution, that is, we provide asymptotic properties for the proposed estimator without the symmetric assumption of the error density. Our empirical studies with simulated data demonstrate that the LMOPG performs well under different distribution of predictor variables with the skewed residual ditribution, and the real data demonstrate that LMOPG has a good fitting.

The rest of paper is organized as follows. In Section 2, we start with our motivation and introduce the detail of the LMOPG method. The effectiveness of the LMOPG algorithm  and the asymptotic properties of LMOPG estimator are derived  in Section 3. A simulation study is conducted in Section 4, which shows the LMOPG performs well under different distribution of predictor variables with the skewed residual ditribution. The Gas Turbine CO and NOx Emission dataset is investigated in section 5 as a real data analysis. We conclude the paper with a short discussion in Section 6. All the proofs are relegated to
Appendix.

\section{Local Modal outer-product gradient method}
\subsection{The Rebuliding Model}
  Suppose $\textsl{y}$ and $\bm{X}$ have joint probability density $f(\textsl{y},\bm{X})$ and conditional probability density $f(\textsl{y}\mid \bm{X})$. The mode of $f(\textsl{y}\mid \bm{X})$ is defined as $\text{argmax}_{\textsl{y}}f(\textsl{y}\mid \bm{X})$, denoted as $\text{Mode}(\textsl{y}\mid \bm{X}).$ The form of MODLR model as follows:
\begin{equation}
\label{MLR}
    \text{Mode}(\textsl{y}\mid \bm{X})=\beta^\top \bm{X}.
\end{equation}
For easy expression, denote $\text{Mode}(\textsl{y}\mid \bm{X})$ as $\text{M}(\textsl{y}\mid \bm{X})$ for the remainder of this paper. Let $\varepsilon=\textsl{y}-\beta^\top \bm{X}$, $g(\varepsilon\mid \bm{X})$ denote the density of $\varepsilon$ given $\bm{X}$. If $g(\varepsilon\mid \bm{X})$ is symmetric about zero, the $\beta$ in \eqref{MLR} will be equivalent to the coefficients obtained by mean linear regression. If $g(\varepsilon\mid \bm{X})$ is skewed, the coefficients obtained from MODLR and mean linear regression will be different. According to this different, MODLR might obtain the information ignored by mean linear regression. 
The following example reveals this. 
\begin{equation*}
\label{Ldiffer}
    \textsl{y}=\beta^\top \bm{X}+\alpha^\top \bm{X}\cdot\varepsilon,
\end{equation*}
where $\bm{X}\perp \!\!\! \perp \varepsilon$ and $\varepsilon$ has skewed density with mean 0 and mode 1. According to MODLR model, we can get $E(\textsl{y}\mid \bm{X})=\beta^\top \bm{X}$ and $\text{M}(\textsl{y}\mid \bm{X})=(\beta+\alpha)^\top \bm{X}$. Moreover,
There are an important example in \citep{li2018sufficient},  that is
\begin{equation}
    \label{Diff}
\textsl{y}=m_{1}\left(\beta_{1}^\top \bm{X}\right)+m_{2}\left(\beta_{2}^\top \bm{X}\right)\cdot \varepsilon,
\end{equation}
where $\varepsilon$ is dependent on $\bm{X}$ with zero mean,  $\beta_{1},\beta_{2} \in \mathbb{R}^{p}$, and $m_{1}$ and $m_{2}$ are unknown link functions. In this case, the SDR method based on the conditional mean $E(\textsl{y}\mid \bm{X})$ can identify one vector $\beta_{1}$ only. By the similar way, we have 
$$\text{M}(\textsl{y}\mid \bm{X})=m_{1}\left(\beta_{1}^\top \bm{X}\right)+m_{2}\left(\beta_{2}^\top \bm{X}\right).$$

Following the MODLR, we can build more general model for handling high dimensional data under model \eqref{M1}. We model the conditional mode of $\textsl{y}$ given $\bm{X}$ as a non-linear function of $\bm{X}$, that is,
\begin{equation}
\label{MNLRg}
\text{M}(\textsl{y}\mid \bm{X})=m(\bm{X})=m\left(\text{B}_{0}^\top \bm{X}\right).
\end{equation}
where $\text{B}_{0}=(\beta_{01},\dots,\beta_{od})$ is the $p\times d$ matrix. 
Furthermore, the new $\varepsilon^{*}$ may be regarded as $\textsl{y}-\text{M}(\textsl{y}\mid \bm{X})$ and its density is skewed. Then, the new model can be constructed as follows:
\begin{equation}
    \label{NewModel}
    \textsl{y}=m\left(\text{B}_0^\top \bm{X}\right)+\varepsilon^{*},
\end{equation}
where $m$ is a unknown function and $\textup{M}\left(\varepsilon^{*}\mid \bm{X}\right)=0$. 
Reconsider example \eqref{Diff}, the $\varepsilon^*$ is $m_2\left(\beta_{2}^\top \bm{X}\right)\times\left(\varepsilon-1\right)$.
So, we can take advantage of model \eqref{MNLRg} to estimate the parameter $\text{B}_{0}$. In particular,  $\text{Span}(\text{B}_{0})$ is special interest. 
\subsection{Methodology}
Suppose the structure dimension is known and $m(\cdot)$ is differentiable, our task is to estimate $\text{Span}(\text{B}_{0})$. 
By the chain rule in \eqref{MNLRg}, we have
\begin{equation}
\label{ChainRul}
    \frac{\partial m(\bm{X})}{\partial \bm{X}}=\frac{\partial m(\text{B}_{0}^\top \bm{X})}{\partial \bm{X}}=\frac{(\partial \text{B}_{0}^\top \bm{X})^\top }{\partial \bm{X}}\frac{\partial m(\text{B}_{0}^\top \bm{X})}{\partial (\text{B}_{0}^\top \bm{X})}=\text{B}_{0}\frac{\partial m(\text{B}_{0}^\top \bm{X})}{\partial (\text{B}_{0}^\top \bm{X})}.
\end{equation}
This shows that the vector $\partial m(\bm{X})/\partial\bm{X}$ belongs to $\text{Span}(\text{B}_{0})$. The following lemma shows that how to obtain the whole basis of $\text{Span}(\text{B}_{0})$ by \eqref{ChainRul}.
\begin{lemma}
\label{Zhidaolem}
    If the function $m(\cdot)$ is differentiable and the $E((\partial m(\bm{X})/\partial \bm{X})(\partial m(\bm{X})/\partial \bm{X}^\top ))$ is a semi-positive defined matrix with rank $d$,
    then 
    $$\textup{Span}\left [ E\left( \frac{\partial m(\bm{X})}{\partial \bm{X}} \frac{\partial m(\bm{X})}{\partial \bm{X}^\top } \right) \right ]=\textup{Span}(\textup{B}_{0}).$$
\end{lemma}

The idea of lemma \ref{Zhidaolem} comes from the OPG method. According to lemma \ref{Zhidaolem}, if we can get the good estimation of $\partial m(\bm{X})/\partial\bm{X}$, then it will be much easier to estimate $\text{Span}(\text{B}_{0}).$ 

Next, we will give an estimation of $\partial m(\bm{X})/\partial\bm{X}$.
Taking a first-order Taylor expansion of $m(\bm{X})$ in \eqref{MNLRg}. For $\bm{X}$ is in the neighbourhood of $X_0$, 
\begin{equation}
\label{TylEpa}
    m(\bm{X})=m(X_0)+(\bm{X}-X_0)^\top  \frac{\partial m(X_0)}{\partial \bm{X}}+\frac{1}{2}(\bm{X}-X_0)^\top \frac{\partial ^{2} m(\bm{X}_{\xi})}{\partial \bm{X}\partial \bm{X}^\top }(\bm{X}-X_0),
\end{equation}
where$\left \|\bm{X}_{\xi}  \right \|_2$ is between $\left \|X_0  \right \|_2$ and $\left \|\bm{X} \right \|_2$, for vector $a$, $\|a\|_2$ stands for L$_{2}$ norm.
Denote
\begin{equation*}
    \frac{\partial m(X_0)}{\partial \bm{X}}=\left( \frac{\partial m(X_0)}{\partial \bm{x}_{1}},\dots,\frac{\partial m(X_0)}{\partial \bm{x}_{p}} \right)^\top \equiv \left(b_{1},\dots,b_{p} \right)^\top =\bm{b}, \ \ m(X_0)\equiv b_{0}.
\end{equation*}
Then,
\begin{equation}
\label{appro}
    m(\bm{X})= b_{0}+\bm{b}^\top (\bm{X}-X_0)+o(\bm{X}-X_0).
\end{equation}

 For fixed $X_0$, the mode of $f(\textsl{y}\mid X_0)$ is equivalent to the mode of $f(\textsl{y}, X_0)$. 
That is 
\begin{equation*}
\label{TransEstimator}
    \mathop{\textup{argmax}}_{\textsl{y}}f(\textsl{y}\mid X_0)= \mathop{\textup{argmax}}_{\textsl{y}}f(\textsl{y}, X_0).
\end{equation*}
 Moreover, the mode of $f(\textsl{y}, X_0)$  is equivalent to the mode of $f(\varepsilon^*, X_0)$ by \eqref{NewModel}. 
 
Suppose $\{(\textsl{y}_{i},\bm{X}_{i})\}_{i=1}^{n}$ is an independent and identically distribution (i.i.d) sample from $f(\textsl{y},\bm{X})$ and \eqref{NewModel}. Then $\hat{f}(\varepsilon^*, X_0)$ at $\varepsilon^*=0$ can be estimated as follows:
\begin{equation*}
    \hat{f}(\varepsilon^*, X_0)=\frac{1}{n}\sum_{i=1}^{n}K_{h_{1}}(\bm{X}_{i}-X_{0})\phi_{h_{2}}\left(\textsl{y}_i-m\left(\text{B}_{0}^\top \bm{X}_i\right)\right),
\end{equation*}
where $K_{h_{1}}$ is a multivariate  kernel function 
\begin{equation*}
    K_{h_{1}}(\bm{X}_{i}-X_{0})=\frac{1}{h_{1_1}h_{1_2}\cdots h_{1_p}}K\left(\frac{\bm{x}_{i1}-x_{01}}{h_{1_1}},\cdots,\frac{\bm{x}_{ip}-x_{0p}}{h_{1_p}}\right),
\end{equation*}
$\phi_{h_{2}}(\textsl{y}_{i}-\textsl{y}_{j})=h_{2}^{-1}\phi ((\textsl{y}_{i}-\textsl{y}_{j})/h_{2})$ is a monomial symmetric kernel function, $h_{1}=(h_{11},\dots,h_{1p})^\top $ and $h_{2}$ are the bandwidths. 
Furthermore, $\textsl{y}_i-m\left(\text{B}_{0}^\top \bm{X}_i\right)$ can be approximated by $\textsl{y}_i-b_{0}-\bm{b}^\top (\bm{X}_i-X_{0})$  by \eqref{appro}.
Let $\theta=(b_{0}, \bm{b}^\top )^\top $ be the parameter to be estimated, it can be solved by maximizing following objective function,
\begin{equation}
\label{Obfunc}
    \text{L}(\theta)\equiv \frac{1}{n}\sum_{i=1}^{n}K_{h_{1}}(\bm{X}_{i}-X_{0})\phi_{h_{2}}\left(\textsl{y}_{i}-b_{0}-\bm{b}^\top (\bm{X}_{i}-X_{0})\right).
\end{equation}
Then, $\hat{\theta}=\text{argmax}_{\theta}\text{L}(\theta)$. 

In practice, if we have $n$ samples. we can give $\bm{X}_j$ in turn, $j=1,\dots,n.$ Then $\hat{\theta}_{(j)}$ relies on giving $\bm{X}_j$, where subscript $(j)$ stands for giving $\bm{X}_j$.
Hence, the basis of $\text{Span}(\text{B}_{0})$ can be estimated by performing spectral decomposition on $n^{-1}\sum_{j=1}^{n}\hat{\bm{b}}_{(j)}\hat{\bm{b}}_{(j)}^\top $ according to lemma \ref{Zhidaolem}, where $\hat{\bm{b}}_{(j)}$ is a $p$-dimensional vector whose elements is the $2nd,\dots,(p+1)$ elements of $\hat{\theta}_{(j)}$.  
Summarizing the above analysis, we can obtain the local modal out-product-gradient (LMOPG) estimator, shown in the Algorithm \ref{LMOPG}.

\floatname{algorithm}{Algorithm}
\begin{algorithm}[H]
\caption{Local Modal Out-Product-Gradient}
\label{LMOPG}
\begin{algorithmic}[1]
\Require $i.i.d.$ sample: $\{\textsl{y}_{i},\bm{X}_{i}\}_{i=1}^{n}$. 
\Ensure Estimators of parameters: $\hat{\theta}_{(j)}$, $j=1,\dots,n$ and the basis of $\text{Span}(\text{B}_{0})$, denoted as $\hat{\text{B}}=(\hat{\beta}_{1},\dots,\hat{\beta}_{d})$.
\State Calculate sample means $\Bar{\bm{X}}$, $\Bar{\textsl{y}}$ and sample variances $\hat{\Sigma}_X$, $\hat{\sigma}_\textsl{y}$ 
\State Calculate $\bm{Z}_i=\hat{\Sigma}_X^{-1/2}(\bm{X}_i-\Bar{\bm{X}})$ and $\tilde{\textsl{y}}_i=\hat{\sigma}_\textsl{y}^{-1/2}(\textsl{y}_i-\Bar{\textsl{y}})$. Let $\bm{Z}_{i}^{*}=(1,\bm{Z}_{i}^\top )^\top $, $i=1,\dots,n$.
\For{$j = 1$ to $n$}
    \State Initialize $\theta^{(0)} = (b_0^{(0)}, b_1^{(0)}, \dots, b_p^{(0)})$ and set $t = 0$
    \Repeat
        \For{$l = 1$ to $n$}
            \State Compute weights 
            $$\qquad \textup{W}(\textit{l}\mid \theta_{(j)}^{(t)})=\frac{K_{h_{1}}(\bm{Z}_{l}-\bm{Z}_{j})\phi_{h_{2}}\left(\tilde{\textsl{y}}_{l}-\theta_{(j)}^{(t)^\top }\left(\bm{Z}_{l}^{*}-\bm{Z}_{j}^{*}\right)\right)}{\sum_{i=1}^{n}K_{h_{1}}(\bm{Z}_{i}-\bm{Z}_{j})\phi_{h_{2}}\left(\tilde{\textsl{y}}_{i}-\theta_{(j)}^{(t)^\top }\left(\bm{Z}_{i}^{*}-\bm{Z}_{j}^{*}\right)\right)}.$$
        \EndFor
        \State Update $\theta_{(j)}^{(t+1)}$ using $$\qquad \theta_{(j)}^{(t+1)}=\mathop{\textup{argmax}}_{\theta_{(j)}}\sum_{l=1}^{n}\left[\textup{W}(\textit{l}\mid \theta_{(j)}^{(t)})\cdot \log\left[\phi_{h_{2}}\left(\tilde{\textsl{y}}_{l}-\theta_{(j)}^\top \left(\bm{Z}_{i}^{*}-\bm{Z}_{j}^{*}\right)\right)\right] \right].$$
        \State $t \gets t + 1$
    \Until{convergence}
\EndFor
 \State Obtain $\hat{\theta}_{(j)}$, $j=1,\dots,n$.
\State  Extract the $2nd,\dots,(p+1)$ elements of $\hat{\theta}_{(j)}$ consists of $\hat{\bm{b}}_{(j)}$, $j=1,\dots,n$.
\State Compute the first $d$ eigenvectors corresponding to the largest $d$ eigenvalues  using spectral decomposition on $1/n\sum_{j=1}^{n} \hat{\bm{b}}_{(j)} \hat{\bm{b}}_{(j)}^\top $, and denoted by $\hat{\nu}_{1},\cdots,\hat{\nu}_{d}$.
\State Obtain $\hat{\text{B}}=(\hat{\beta}_{1},\dots,\hat{\beta}_{d})=\hat{\Sigma}_X^{-1/2}(\hat{\nu}_{1},\cdots,\hat{\nu}_{d})$. 
\end{algorithmic} 
\end{algorithm}

Although the choice of kernel function is not very important, the aim of standardizing $\bm{X}_1,\dots,\bm{X}_n$ is to guarantee that all the predictors are on the same scale. It is helpful to use a spherically-shaped kernel function and choose the same bandwidth for each component of $\bm{X}$ in practice and we use Gaussian kernel for $\phi_{h_{2}}$ in the rest of paper.
In particularly, the optimal solution of step {\footnotesize 9} in Algorithm \ref{LMOPG} is $\left(\mathbb{Z}^\top \mathbb{W}^{(t)}\mathbb{Z}\right)^{-1}\mathbb{Z}^\top \mathbb{W}^{(t)}\bm{\text{Y}}$ when using Gaussian kernel, where $\mathbb{Z}=(\bm{Z}_{1}^{*},\dots,\bm{Z}_{n}^{*})^\top $ is a $n\times (p+1)$ matrix, $\mathbb{W}^{(t)}$ is a $n\times n$ diagonal matrix with diagonal elements $\textup{W}(\textit{l}\mid \theta_{(j)}^{(t)})$ and $\bm{\text{Y}}=(\tilde{\textsl{y}}_{1},\dots,\tilde{\textsl{y}}_{n})^\top $ is a $n$-dimensional vector.
Moreover,  notice that the parameter $\theta$ we want to estimate is in the one-dimensional kernel function $\phi_{h_{2}}$ and use an algorithm similar to the EM algorithm, each maximization of the objective function is done in a weighted one-dimensional kernel function. So, although the high dimensional kernel function $K_{h_{1}}$ contained in $\text{L}(\theta)$, the proposed LMOPG in section \ref{Simu} exhibits promising performance.

\section{Theoretical Properties}
In this section, we will prove the validity of the algorithm and $\{\hat{\beta}_{i}\}_{i=1}^{k}$ converge to  a set of standard orthogonal basis for $\textup{Span}(\textup{B}_{0})$ under the mild condition.  
\begin{theorem}
\label{iterTh}
    Each iteration of step {\footnotesize 2} and step {\footnotesize 3} in Algorithm \ref{LMOPG} will monotonically non-decrease the objective
function \ref{Obfunc}, that is, $\text{L}(\theta^{(t+1)})\geq \text{L}(\theta^{(t)})$, for all $k$.
\end{theorem}
Before giving the asymptotic properties of $\hat{\text{B}}$, we need to give some mild conditions and notations.  

(A1) The $m(\bm{X})$ is continuous $2$-th derivative.

(A2) The condition density function $g(\varepsilon^{*}\mid \bm{X})$ satisfies $g'(0\mid\bm{X})=0$, $g''(0\mid\bm{X})<0$, $g^{v}(t\mid\bm{X})$ is bounded  in a neighbor of $X_0$ and has continuous first derivative at the point $X_0$ as a function of $\bm{X}$, for $v=0,\dots,2$.

(A3) The density function $f(\bm{X})$ is bounded and has continuous first derivative at the point $X_0$ and $f(X_0)>0$.

(A4) $K(\cdot)$ is a kernel density function with a bounded derivative and support. All the moments of $K(\cdot)$ exist.

(A5) $\bm{X}$ is uniformly integrable in the neighbor of  $X_0$.


Next, we define some notations. For easy presentation in the proofing process, let $h_{1}=h_{1_1}=\cdots=h_{1_p}$. Define $H=diag\{1,h_{1},\dots,h_{1}\}$ is a $(p+1)\times(p+1)$ diagonal matrix, $\theta=(b_{0},b_{1},\dots,b_{p})^\top $, $\theta^{*}=H\theta$, $$R(\bm{X}_{i})=m(\bm{X}_{i})-m(X_{0})-\frac{\partial m(X_0)}{\partial \bm{X}_{i}^\top }(\bm{X}_{i}-X_0)=m(\bm{X}_{i})-b_{0}-\bm{b}^\top (\bm{X}_{i}-X_0)$$
and $\bm{\text{X}}_{i}^{*}=\left(1,(\bm{x}_{i1}-x_{01})/h_{1},\dots,(\bm{x}_{ip}-x_{0p})/h_{1}\right)^\top .$ So, $\textsl{y}_{i}-b_{0}-\bm{b}^\top (\bm{X}_{i}-X_0)=\varepsilon_{i}^{*}+R(\bm{X}_{i})$.

\begin{theorem}
\label{consistentTheta}
    Under the regularity conditions (A1)-(A4), if the bandwidths $h_{1}$ and $h_{2}$ go to zero (have the same order) such that $nh_{1}^{p}h_{2}^{5}\rightarrow \infty $, $h_2^2/h_1\rightarrow 0$, and $h_1^2/h_2\rightarrow 0$, 
    there exists a consistent local maximizer $\hat{\theta}$ of \eqref{Obfunc} such that 
$$\left\|\hat{\theta}^*-\theta^*\right\|_2=O_{p}\left((nh_{1}^ph_{2}^{3})^{-1/2}+h_{1}^{2}+h_{2}^{2}\right),$$
   Furthermore, let $E(\bm{b}\bm{b}^\top )=\textup{B}$. If $\|\hat{\bm{b}}\|_2$, $\|\bm{b}\|_2$, $\|\hat{\textup{B}}\|_2$ and $\|\textup{B}\|_2$ are finite, then
$$h_1\left\|\hat{\textup{B}}\hat{\textup{B}}^\top -\textup{B}_0\textup{B}_0^\top \right\|_F=O_{p}\left((nh_{1}^ph_{2}^{3})^{-1/2}+h_{1}^{2}+h_{2}^{2}\right),$$
 where $\hat{\textup{B}}$ is the estimate of ~ $\textup{B}$. For matrix $A$, $\|A\|_2$ stands for  spectral norm, $\|A\|_F$ stands for  Frobenius norm.
\end{theorem}

\begin{theorem}
   Let $\bm{t}=\left( t_1,\cdots,t_p\right)^\top $. Under the regularity conditions (A1)-(A4),  if the bandwidths $h_{1}$ and $h_{2}$ go to zero such that $nh_{1}^{p}h_{2}^{5}\rightarrow \infty $,  then
   \begin{equation}
        \left[\textup{Var}\left(\hat{\theta}^*\right)\right]^{-\frac{1}{2}}\left(\hat{\theta}^*+\frac{h_2^2g'''(0\mid X_0)\tilde{\Delta}_1^{-1}\upsilon_{\textup{I}_1}}{2g''(0\mid X_{0})}-\tilde{\mathbf{b}} \right)\stackrel{L}{\longrightarrow} \mathrm{N}(0, I_{p+1}),
   \end{equation}
   where $$\tilde{\mathbf{b}}\approx \frac{h_1^2\tilde{\Delta}_1^{-1}\upsilon_{\textup{I}_2}}{2}-\frac{h_2^2g'''(0\mid X_0)\tilde{\Delta}_1^{-1}\upsilon_{\textup{I}_1}}{2g''(0\mid X_{0})},$$
   $$\tilde{\Delta}_1=\int\cdots\int K(t_1,\dots,t_p)(1,\bm{t}^\top )^\top (1,\bm{t}^\top )\text{d}t_1\cdots \text{d}t_p,$$ 
   $$\upsilon_{\textup{I}_1}=\int\cdots\int K\left( t_1,\cdots,t_p\right)(1,\bm{t}^\top )^\top \text{d}t_1\cdots \text{d}t_p,$$ and

$$\upsilon_{\textup{I}_2}=\int \cdots \int K\left( t_1,\cdots,t_p\right)(1,\bm{t}^\top )^\top  \bm{t}^Tm^{(2)}(X_{0} )\bm{t}
\text{d}t_1\cdots \text{d}t_p,$$
$m^{(2)}(X_{0} )$ is the second-order partial derivative at $X_0$ with respect to $\bm{X}$.
\end{theorem}

\section{Simulation studies}
\label{Simu}

In this section, we evaluate the performance of the proposed LMOPG method by using a Monte Carlo simulation study. To measure the distance between the true subspace $\textup{Span}(\textup{B}_{0})$ and the corresponding estimator $\textup{Span}(\hat{\text{B}})$ for $\hat{\text{B}}=(\hat{\beta_{1}}, \ldots,\hat{\beta_{d}})$, we consider the trace correlation defined as  \citep{ferre1998determining,dong2015robust} 
\begin{equation}
\label{eq14}
R   = \dfrac{\mathrm{trace}\left (P_{\text{B}_{0}}P_{\hat{\text{B}}}\right )}{d},
\end{equation}
where $P_{\text{B}_{0}}=\text{B}_{0}(\text{B}_{0}^\top\text{B}_{0})^{-1}\text{B}_{0}^\top$ denotes the projection matrix. 
Without loss of generality, we assume $\hat{\text{B}}$ is a column-orthogonal matrix due to the property of $\text{B}_{0}$. Otherwise, the Gram-Schmidt ortho-normalization method can be used and will not change the subspace. Then, the trace correlation based on the estimator $\hat{\text{B}}$ in Eq. \eqref{eq14} can be calculated as
\begin{equation}
R  =\dfrac{\mathrm{trace}\left (\hat{\text{B}}^\top \textup{B}_{0}\textup{B}_{0}^\top\hat{\text{B}}\right )}{d}.
\end{equation}
Here, the trace correlation $R$ can be used to evaluate and compare the performance of different estimation methods. The trace correlation is a value between 0 and 1, and a larger value of $R$ indicates a better estimator $\hat{\text{B}}$.  In the following subsections, we consider $\bm{X}$ follows Normal distributions in Section \ref{sec4.1} to investigate the validity of the proposed LMOPG method, and $\bm{X}$ follows non-Normal distributions in Section \ref{sec4.2} to evaluate the robustness of different distribution. 

\subsection{Normal distribution}
\label{sec4.1}

We  consider comparing the LMOPG method to the SIR, SAVE, and PHD with residuals version methods under five different models with normally distributed predictor variables studied or motivated in Example 1.(i) of \cite{yao2014new}, Example 1 of \cite{xia2002adaptive}, Example 9.1 of \cite{li2018sufficient}, Example 3 of \cite{Li1991}, and Example 1 of \cite{li2007directional}:

\begin{itemize}
    \item {\bf Model [A1]}:
\begin{equation*}
    \textsl{y}=\beta_{1}^\top \bm{X}+\beta_{2}^\top \bm{X}\times \varepsilon
\end{equation*}

   \item {\bf Model [A2]}:
\begin{equation*}
	\textsl{y}=2\mathrm{sin}\left(1.4 \beta_{1}^\top\bm{X}\right)+\left( \beta_{2}^\top\bm{X}+1\right)^{2}\times\varepsilon.
\end{equation*}

   \item {\bf Model [A3]}:
\begin{equation*}
\textsl{y}=\beta_{1}^\top\bm{X}/\left\{0.5+(\beta_{2}^\top\bm{X}+1.5)^{2}\right\}+\sigma\varepsilon. 
\end{equation*}

   \item {\bf Model [A4]}:
\begin{equation*}
\textsl{y}=\beta_{1}^\top\bm{X}\times (\beta_{2}^\top\bm{X}+1)+\sigma\varepsilon. 
\end{equation*}

   \item {\bf Model [A5]}:
\begin{equation*}
\textsl{y}=0.4(\beta_{1}^\top\bm{X})+3\sin\{(\beta_{1}^\top\bm{X}\times \beta_{2}^\top\bm{X})/4\} +\sigma\varepsilon. 
\end{equation*}

\end{itemize}

\begin{figure}[ht]
\centering
\includegraphics[width=0.7\textwidth]{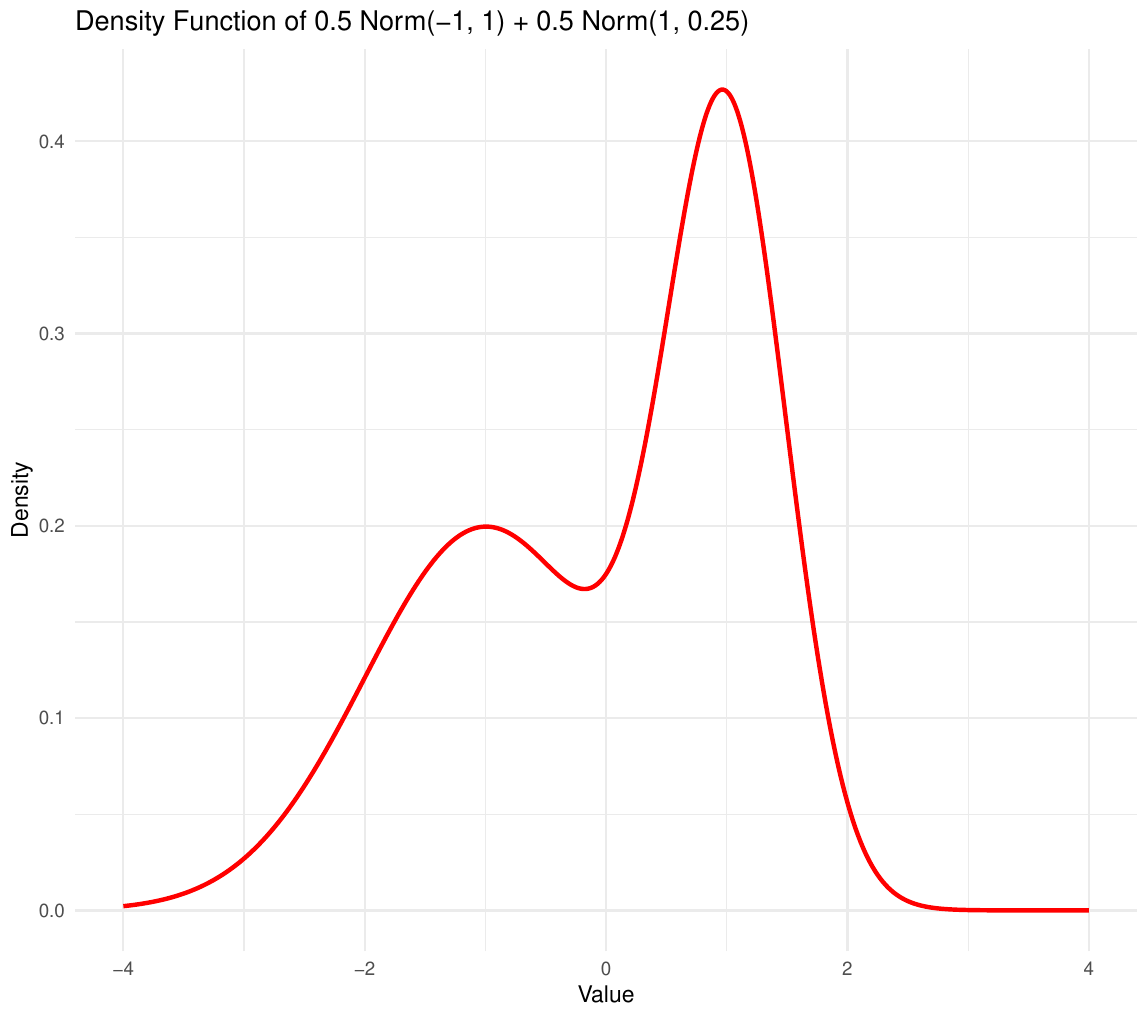}
    \caption{The density function of $0.5Norm(-1,1)+0.5Norm(1,0.25)$.}
\label{Epsilon.density}
\end{figure}

We set $\mathrm{dim}(\bm{X})=10$, $\bm{X}\sim N_{10}(\bm{0}, \Sigma)$,  $\varepsilon\sim 0.5N(-1,1)+0.5N(1,0.25)$ and its density function is shown as Figure \ref{Epsilon.density},  $\sigma=0.5$, $\beta_{1}=(1,0,\dots,0)$ and $\beta_{2}=(0,1,\dots,0)$, where $\Sigma$ is a $p\times p$ diagonal matrix, the number of slices $H$ to be 10, and the sample sizes are $n = 200$, $300$ and $500$. We use multivariate Gaussian kernel for $K_{h_{1}}$ and $K_{h_{2}}$ and chose constant $1$ as the bandwidth for both $h_{1}$ and $h_{2}$ in this subsection. In this paper, we do not give the way of choosing the optimal bandwidth, because it is hard to estimated the second and higher derivatives of link function $g$.  In the following simulations we find that LMOPG performs well even though the bandwidth is not optimal.

The SIR, SAVE and PHD methods are implemented in the R packages {\tt dr}  \citep{drpackage2002}.
The averages and standard deviations (SDs) of $R$ for the proposed LMOPG method, the SIR method, the SAVE method, and the PHD method based on 100 simulations are reported in Table \ref{Normal}. 

From the results in table \ref{Normal},  the performance of the methods considered here improved with the increase in sample size. We consider simulations starting with a sample size of 200, because the multivariate kernel function requires a large sample size and the convergence of the SAVE method is slower.
the We find that the PHDres method perform  poorly in model {\bf [A1]} and {\bf [A2]} and perform well in moder {\bf [A3]-[A5]}. This is consistent with the theoretical results because the PHDres is densigend to estimate the central mean subspace.  The LMOPG method performs better than the other methods  in almost all cases considered here, which indicates that the LMOPG method is an effective method for estimating $\mathrm{Span}(\text{B})$ in the multivariate normal case, skewed residuals distribution.

\begin{table}[!p]
\centering
	\caption{
Averages and standard deviations of the trace correlation $R$ for the LMOPG method for model {\bf [A1]}, {\bf [A2]}, {\bf [A3]}, {\bf [A4]} and {\bf [A5]} with multivariate normally distributed predicted variables based on 1000 simulations with different sample sizes.}
\label{Normal}
        \setlength{\tabcolsep}{10.5pt} 
        \renewcommand{\arraystretch}{1.1}
\begin{tabular}{cccclll}
\hline
\multirow{2}{*}{Normal} &
  \multicolumn{2}{c}{Model} &
  \multicolumn{4}{c}{A1} \\
 &
  \multicolumn{2}{c}{Method} &
  LMOPG &
  \multicolumn{1}{c}{SIR} &
  \multicolumn{1}{c}{SAVE} &
  \multicolumn{1}{c}{PHDres} \\ \hline
\multicolumn{2}{c}{$n=200$} &
  \begin{tabular}[c]{@{}c@{}}Average\\ SD\end{tabular} &
  \begin{tabular}[c]{@{}c@{}}0.7936\\ 0.1272\end{tabular} &
  \multicolumn{1}{c}{\begin{tabular}[c]{@{}c@{}}0.5585\\ 0.0796\end{tabular}} &
  \multicolumn{1}{c}{\begin{tabular}[c]{@{}c@{}}0.5338\\ 0.1249\end{tabular}} &
  \multicolumn{1}{c}{\begin{tabular}[c]{@{}c@{}}0.3436\\ 0.1223\end{tabular}} \\
\multicolumn{2}{c}{$n=300$} &
  \begin{tabular}[c]{@{}c@{}}Average\\ SD\end{tabular} &
  \begin{tabular}[c]{@{}c@{}}0.8838\\ 0.0658\end{tabular} &
  \multicolumn{1}{c}{\begin{tabular}[c]{@{}c@{}}0.5544\\ 0.0657\end{tabular}} &
  \multicolumn{1}{c}{\begin{tabular}[c]{@{}c@{}}0.6152\\ 0.1140\end{tabular}} &
  \multicolumn{1}{c}{\begin{tabular}[c]{@{}c@{}}0.3422\\ 0.1193\end{tabular}} \\
\multicolumn{2}{c}{$n=500$} &
  \begin{tabular}[c]{@{}c@{}}Average\\ SD\end{tabular} &
  \begin{tabular}[c]{@{}c@{}}0.9372\\ 0.0259\end{tabular} &
  \multicolumn{1}{c}{\begin{tabular}[c]{@{}c@{}}0.5459\\ 0.0630\end{tabular}} &
  \multicolumn{1}{c}{\begin{tabular}[c]{@{}c@{}}0.6346\\ 0.1242\end{tabular}} &
  \multicolumn{1}{c}{\begin{tabular}[c]{@{}c@{}}0.3513\\ 0.1251\end{tabular}} \\ \hline
 &
  \multicolumn{2}{c}{Model} &
  \multicolumn{4}{c}{A2} \\
\multicolumn{2}{c}{$n=200$} &
  \begin{tabular}[c]{@{}c@{}}Average\\ SD\end{tabular} &
  \begin{tabular}[c]{@{}c@{}}0.7827\\ 0.1287\end{tabular} &
  \multicolumn{1}{c}{\begin{tabular}[c]{@{}c@{}}0.5464\\ 0.0759\end{tabular}} &
  \multicolumn{1}{c}{\begin{tabular}[c]{@{}c@{}}0.4480\\ 0.1634\end{tabular}} &
  \multicolumn{1}{c}{\begin{tabular}[c]{@{}c@{}}0.4123\\ 0.1169\end{tabular}} \\
\multicolumn{2}{c}{$n=300$} &
  \begin{tabular}[c]{@{}c@{}}Average\\ SD\end{tabular} &
  \begin{tabular}[c]{@{}c@{}}0.8707\\ 0.0828\end{tabular} &
  \multicolumn{1}{c}{\begin{tabular}[c]{@{}c@{}}0.5417\\ 0.0652\end{tabular}} &
  \multicolumn{1}{c}{\begin{tabular}[c]{@{}c@{}}0.6015\\ 0.1683\end{tabular}} &
  \multicolumn{1}{c}{\begin{tabular}[c]{@{}c@{}}0.4450\\ 0.1473\end{tabular}} \\
\multicolumn{2}{c}{$n=500$} &
  \begin{tabular}[c]{@{}c@{}}Average\\ SD\end{tabular} &
  \begin{tabular}[c]{@{}c@{}}0.9358\\ 0.0478\end{tabular} &
  \multicolumn{1}{c}{\begin{tabular}[c]{@{}c@{}}0.5571\\ 0.0828\end{tabular}} &
  \multicolumn{1}{c}{\begin{tabular}[c]{@{}c@{}}0.8086\\ 0.1291\end{tabular}} &
  \multicolumn{1}{c}{\begin{tabular}[c]{@{}c@{}}0.4702\\ 0.1343\end{tabular}} \\ \hline
 &
  \multicolumn{2}{c}{Model} &
  \multicolumn{4}{c}{A3} \\
\multicolumn{2}{c}{$n=200$} &
  \begin{tabular}[c]{@{}c@{}}Average\\ SD\end{tabular} &
  \multicolumn{1}{c}{\begin{tabular}[c]{@{}c@{}}0.9274\\ 0.0371\end{tabular}} &
  \begin{tabular}[c]{@{}l@{}}0.9205\\ 0.0321\end{tabular} &
  \begin{tabular}[c]{@{}l@{}}0.2548\\ 0.1461\end{tabular} &
  \begin{tabular}[c]{@{}l@{}}0.8751\\ 0.0432\end{tabular} \\
\multicolumn{2}{c}{$n=300$} &
  \begin{tabular}[c]{@{}c@{}}Average\\ SD\end{tabular} &
  \multicolumn{1}{c}{\begin{tabular}[c]{@{}c@{}}0.9555\\ 0.0215\end{tabular}} &
  \begin{tabular}[c]{@{}l@{}}0.9524\\ 0.0199\end{tabular} &
  \begin{tabular}[c]{@{}l@{}}0.4707\\ 0.1455\end{tabular} &
  \begin{tabular}[c]{@{}l@{}}0.9249\\ 0.0340\end{tabular} \\
\multicolumn{2}{c}{$n=500$} &
  \begin{tabular}[c]{@{}c@{}}Average\\ SD\end{tabular} &
  \multicolumn{1}{c}{\begin{tabular}[c]{@{}c@{}}0.9703\\ 0.0131\end{tabular}} &
  \begin{tabular}[c]{@{}l@{}}0.9703\\ 0.0131\end{tabular} &
  \begin{tabular}[c]{@{}l@{}}0.6648\\ 0.1391\end{tabular} &
  \begin{tabular}[c]{@{}l@{}}0.9530\\ 0.0185\end{tabular} \\ \hline
\multicolumn{1}{l}{} &
  \multicolumn{2}{c}{Model} &
  \multicolumn{4}{c}{A4} \\
\multicolumn{2}{c}{$n=200$} &
  \begin{tabular}[c]{@{}c@{}}Average\\ SD\end{tabular} &
  \multicolumn{1}{c}{\begin{tabular}[c]{@{}c@{}}0.9740\\ 0.0096\end{tabular}} &
  \begin{tabular}[c]{@{}l@{}}0.8914\\ 0.0619\end{tabular} &
  \begin{tabular}[c]{@{}l@{}}0.3731\\ 0.1546\end{tabular} &
  \begin{tabular}[c]{@{}l@{}}0.9150\\ 0.0361\end{tabular} \\
\multicolumn{2}{c}{$n=300$} &
  \begin{tabular}[c]{@{}c@{}}Average\\ SD\end{tabular} &
  \multicolumn{1}{c}{\begin{tabular}[c]{@{}c@{}}0.9849\\ 0.0051\end{tabular}} &
  \begin{tabular}[c]{@{}l@{}}0.9309\\ 0.0360\end{tabular} &
  \begin{tabular}[c]{@{}l@{}}0.5091\\ 0.1479\end{tabular} &
  \begin{tabular}[c]{@{}l@{}}0.9479\\ 0.0219\end{tabular} \\
\multicolumn{2}{c}{$n=500$} &
  \begin{tabular}[c]{@{}c@{}}Average\\ SD\end{tabular} &
  \multicolumn{1}{c}{\begin{tabular}[c]{@{}c@{}}0.9913\\ 0.0032\end{tabular}} &
  \begin{tabular}[c]{@{}l@{}}0.9591\\ 0.0172\end{tabular} &
  \begin{tabular}[c]{@{}l@{}}0.6998\\ 0.1468\end{tabular} &
  \begin{tabular}[c]{@{}l@{}}0.9627\\ 0.0141\end{tabular} \\ \hline
\multicolumn{1}{l}{} &
  \multicolumn{2}{c}{Model} &
  \multicolumn{4}{c}{A5} \\
\multicolumn{2}{c}{$n=200$} &
  \begin{tabular}[c]{@{}c@{}}Average\\ SD\end{tabular} &
  \multicolumn{1}{c}{\begin{tabular}[c]{@{}c@{}}0.9577\\ 0.0165\end{tabular}} &
  \begin{tabular}[c]{@{}l@{}}0.6635\\ 0.1478\end{tabular} &
  \begin{tabular}[c]{@{}l@{}}0.6072\\ 0.1968\end{tabular} &
  \begin{tabular}[c]{@{}l@{}}0.9213\\ 0.0320\end{tabular} \\
\multicolumn{2}{c}{$n=300$} &
  \begin{tabular}[c]{@{}c@{}}Average\\ SD\end{tabular} &
  \multicolumn{1}{c}{\begin{tabular}[c]{@{}c@{}}0.9761\\ 0.0083\end{tabular}} &
  \begin{tabular}[c]{@{}l@{}}0.7666\\ 0.1225\end{tabular} &
  \begin{tabular}[c]{@{}l@{}}0.8117\\ 0.1170\end{tabular} &
  \begin{tabular}[c]{@{}l@{}}0.9505\\ 0.0201\end{tabular} \\
\multicolumn{2}{c}{$n=500$} &
  \begin{tabular}[c]{@{}c@{}}Average\\ SD\end{tabular} &
  \multicolumn{1}{c}{\begin{tabular}[c]{@{}c@{}}0.9868\\ 0.0047\end{tabular}} &
  \begin{tabular}[c]{@{}l@{}}0.8759\\ 0.0549\end{tabular} &
  \begin{tabular}[c]{@{}l@{}}0.9177\\ 0.0373\end{tabular} &
  \begin{tabular}[c]{@{}l@{}}0.9660\\ 0.0124\end{tabular} \\ \hline
\end{tabular}
\end{table}

\subsection{Non-Normal distribution}
\label{sec4.2}
\begin{figure}[ht]
\centering
\includegraphics[width=0.7\textwidth]{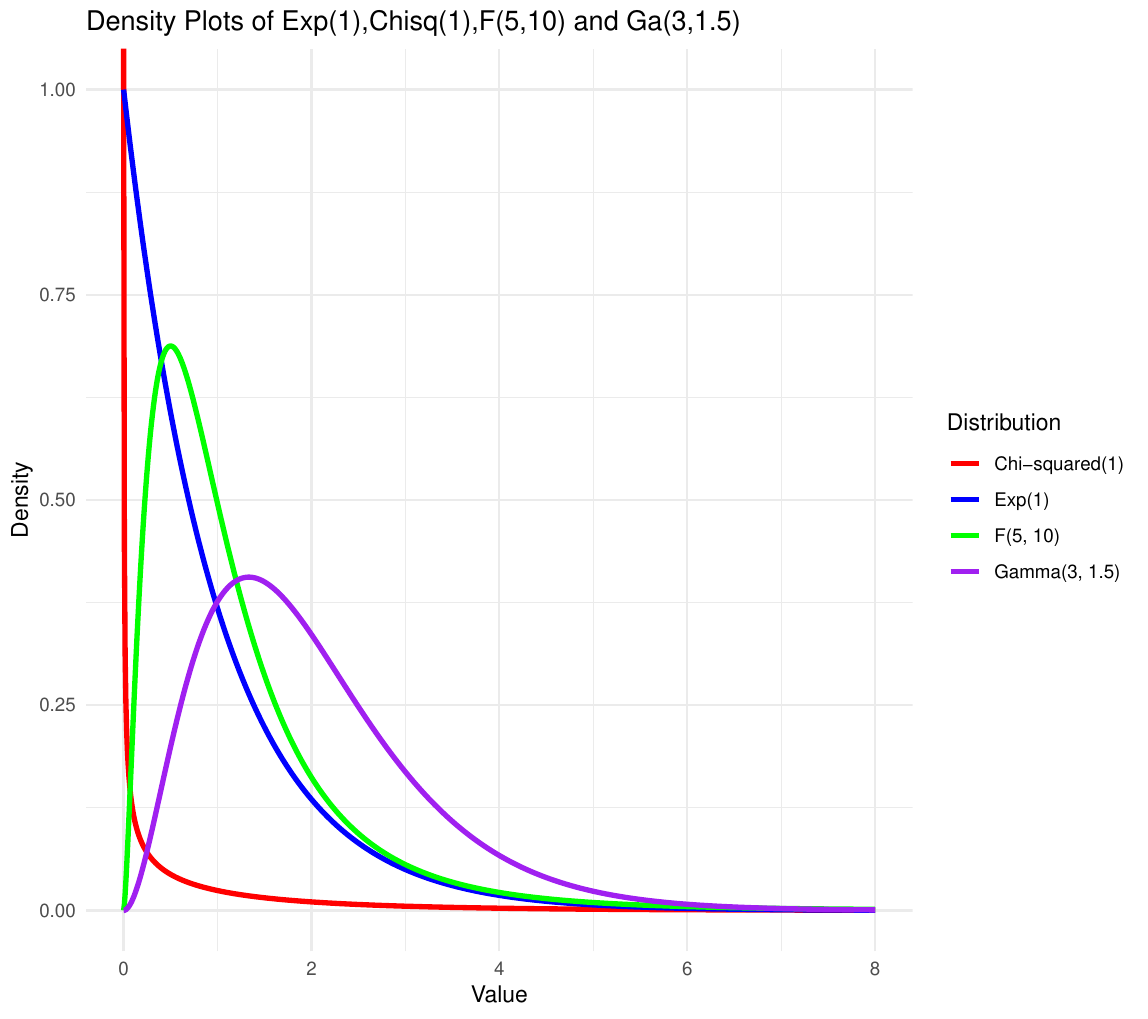}
    \caption{The density function of Exp(1), Chisq(1), F(5,10) and Ga(3,1.5) about predictor variable.}
\label{density}
\end{figure}

In this subsection, we consider the predictor variables following different non-normal multivariate elliptical distributions. Specifically, the multivariate Chi-squares with degree 1, multivariate Exponential with rate parameter 1, multivariate F with degrees 5 and 10, and multivariate Gamma with shape paremeter 3 and scale parameter 1.5 distributions are considered. The above four univariate  different density functions are shown as Figure \ref{density}. From Figure \ref{density}, we can see that the values of the random variable become more dispersed. So, We chose constant 7 as the bandwidth under the multivariate Chi-squares and chose 8 as the bandwidth under others similarly. The reason we make the bandwidth larger is the multivariate kernel function requires a larger sample size. 
Since the above four distributions are not symmetric, we compare the proposed LMOPG with those methods that do not strictly restrict the distribution of the predictor variables, such as MAVE and OPG for the central mean subspace (denoted as meanMAVE and meanOPG respectively) and MAVE and OPG for the central subspace (denoted as csMAVE and csOPG respectively). The OPG and MAVE methods are implemented in the R package {\tt MAVE} \citep{MAVE2021}. 

Furthermore, the following three models are considered. Models {\bf [B1]} and  {\bf [B2]} are based on Example 2 of \cite{zhu2006sliced} with a slow-growing power function of degree $1/2$. The model {\bf [B3]} is is motivated by Example 1 of \cite{li2007directional}. Finally, we compare the proposed LMOPG with the meanOPG, the csOPG, the meanMAVE and the csMAVE under four different distribution of predictor variables, models {\bf [A1]} and {\bf [B1]-[B3]}.  Table \ref{Chisq}-\ref{Gamma} report the averages and standard deviations of the trace correlation $R$ based on 100 simulations.

\begin{itemize}
    \item {\bf Model [B1]}:
\begin{equation*}
   \textsl{y}=|(4+ \beta_{1}^\top\bm{X})|^{1/2}\times |(2+ \beta_{2}^\top\bm{X})|^{1/2}+\sigma\varepsilon.
\end{equation*}

    \item {\bf Model [B2]}:
\begin{equation*}
   \textsl{y}=|\beta_{1}^\top\bm{X}|^{1/2}+ | \beta_{2}^\top\bm{X}\times\varepsilon|^{1/2}+\sigma\varepsilon.
\end{equation*}

    \item {\bf Model [B3]}:
\begin{equation*}
Y=0.4\left(\beta_{1}^\top\bm{X}\right)+3\mathrm{sin}\left ( \beta_{2}^\top\bm{X}/4 \right )+\sigma\varepsilon.
\end{equation*}
\end{itemize}

\begin{table}[]
\caption{
Averages and standard deviations of the trace correlation $R$ for the LMOPG method for model {\bf [A1]}, {\bf [B1]}, {\bf [B2]} and {\bf [B3]} with multivariate Chi-squared distributed predicted variables based on 100 simulations with different sample sizes.}
\label{Chisq}
        \setlength{\tabcolsep}{5.5pt} 
        \renewcommand{\arraystretch}{1.1}
        \hspace*{0.6em}
\begin{tabular}{cccccccc}
\hline
\multicolumn{1}{c}{\multirow{2}{*}{ Chis-q ~}} &
  \multicolumn{2}{c}{Model} &
  \multicolumn{5}{c}{A1} \\
\multicolumn{1}{c}{} &
  \multicolumn{2}{c}{Method} &
  LMOPG &meanOPG &csOPG &meanMAVE &csMAVE \\ \hline
\multicolumn{2}{c}{$n=200$} &
  \begin{tabular}[c]{@{}c@{}}Average\\ SD\end{tabular} &
  \begin{tabular}[c]{@{}c@{}}0.8542\\ 0.0896\end{tabular} &
  \begin{tabular}[c]{@{}c@{}}0.6070\\ 0.1152\end{tabular} &
  \begin{tabular}[c]{@{}c@{}}0.8013\\ 0.1559\end{tabular} &
  \begin{tabular}[c]{@{}c@{}}0.6095\\ 0.0858\end{tabular} &
  \begin{tabular}[c]{@{}c@{}}0.8996\\ 0.1070\end{tabular} \\
\multicolumn{2}{c}{$n=300$} &
  \begin{tabular}[c]{@{}c@{}}Average\\ SD\end{tabular} &
  \begin{tabular}[c]{@{}c@{}}0.8952\\ 0.0734\end{tabular} &
  \begin{tabular}[c]{@{}c@{}}0.6138\\ 0.1207\end{tabular} &
  \begin{tabular}[c]{@{}c@{}}0.8703\\ 0.1386\end{tabular} &
  \begin{tabular}[c]{@{}c@{}}0.6216\\ 0.0875\end{tabular} &
  \begin{tabular}[c]{@{}c@{}}0.9499\\ 0.0639\end{tabular} \\
\multicolumn{2}{c}{$n=500$} &
  \begin{tabular}[c]{@{}c@{}}Average\\ SD\end{tabular} &
  \begin{tabular}[c]{@{}c@{}}0.9255\\ 0.0587\end{tabular} &
  \begin{tabular}[c]{@{}c@{}}0.6024\\ 0.1133\end{tabular} &
  \begin{tabular}[c]{@{}c@{}}0.9232\\ 0.1043\end{tabular} &
  \begin{tabular}[c]{@{}c@{}}0.6178\\ 0.0866\end{tabular} &
  \begin{tabular}[c]{@{}c@{}}0.9785\\ 0.0364\end{tabular} \\ \hline
 &
  \multicolumn{2}{c}{Model} &
  \multicolumn{5}{c}{B1} \\
\multicolumn{2}{c}{$n=200$} &
  \begin{tabular}[c]{@{}c@{}}Average\\ SD\end{tabular} &
  \begin{tabular}[c]{@{}c@{}}0.7432\\ 0.1285\end{tabular} &
  \begin{tabular}[c]{@{}c@{}}0.7627\\ 0.1639\end{tabular} &
  \begin{tabular}[c]{@{}c@{}}0.6706\\ 0.1706\end{tabular} &
  \begin{tabular}[c]{@{}c@{}}0.7944\\ 0.1446\end{tabular} &
  \begin{tabular}[c]{@{}c@{}}0.5940\\ 0.1127\end{tabular} \\
\multicolumn{2}{c}{$n=300$} &
  \begin{tabular}[c]{@{}c@{}}Average\\ SD\end{tabular} &
  \begin{tabular}[c]{@{}c@{}}0.8061\\ 0.1304\end{tabular} &
  \begin{tabular}[c]{@{}c@{}}0.8392\\ 0.1535\end{tabular} &
  \begin{tabular}[c]{@{}c@{}}0.6934\\ 0.1936\end{tabular} &
  \begin{tabular}[c]{@{}c@{}}0.8656\\ 0.1194\end{tabular} &
  \begin{tabular}[c]{@{}c@{}}0.6512\\ 0.1507\end{tabular} \\
\multicolumn{2}{c}{$n=500$} &
  \begin{tabular}[c]{@{}c@{}}Average\\ SD\end{tabular} &
  \begin{tabular}[c]{@{}c@{}}0.8894\\ 0.0796\end{tabular} &
  \begin{tabular}[c]{@{}c@{}}0.8776\\ 0.1279\end{tabular} &
  \begin{tabular}[c]{@{}c@{}}0.8593\\ 0.1695\end{tabular} &
  \begin{tabular}[c]{@{}c@{}}0.9229\\ 0.0848\end{tabular} &
  \begin{tabular}[c]{@{}c@{}}0.7270\\ 0.1647\end{tabular} \\ \hline
  &
  \multicolumn{2}{c}{Model} &
  \multicolumn{5}{c}{B2} \\
\multicolumn{2}{c}{$n=200$} &
  \begin{tabular}[c]{@{}c@{}}Average\\ SD\end{tabular} &
  \begin{tabular}[c]{@{}c@{}}0.7445\\ 0.1512\end{tabular} &
  \begin{tabular}[c]{@{}c@{}}0.6989\\ 0.1760\end{tabular} &
  \begin{tabular}[c]{@{}c@{}}0.7134\\ 0.1764\end{tabular} &
  \begin{tabular}[c]{@{}c@{}}0.7600\\ 0.1723\end{tabular} &
  \begin{tabular}[c]{@{}c@{}}0.6829\\ 0.1511\end{tabular} \\
\multicolumn{2}{c}{$n=300$} &
  \begin{tabular}[c]{@{}c@{}}Average\\ SD\end{tabular} &
  \begin{tabular}[c]{@{}c@{}}0.8439\\ 0.1157\end{tabular} &
  \begin{tabular}[c]{@{}c@{}}0.7247\\ 0.1729\end{tabular} &
  \begin{tabular}[c]{@{}c@{}}0.7705\\ 0.1867\end{tabular} &
  \begin{tabular}[c]{@{}c@{}}0.8279\\ 0.1624\end{tabular} &
  \begin{tabular}[c]{@{}c@{}}0.7345\\ 0.1618\end{tabular} \\
\multicolumn{2}{c}{$n=500$} &
  \begin{tabular}[c]{@{}c@{}}Average\\ SD\end{tabular} &
  \begin{tabular}[c]{@{}c@{}}0.8925\\ 0.0991\end{tabular} &
  \begin{tabular}[c]{@{}c@{}}0.7675\\ 0.1892\end{tabular} &
  \begin{tabular}[c]{@{}c@{}}0.8473\\ 0.1714\end{tabular} &
  \begin{tabular}[c]{@{}c@{}}0.9124\\ 0.1191\end{tabular} &
  \begin{tabular}[c]{@{}c@{}}0.8605\\ 0.1511\end{tabular} \\ \hline
 &
  \multicolumn{2}{c}{Model} &
  \multicolumn{5}{c}{B3} \\
\multicolumn{2}{c}{$n=200$} &
  \begin{tabular}[c]{@{}c@{}}Average\\ SD\end{tabular} &
  \begin{tabular}[c]{@{}c@{}}0.7235\\ 0.1270\end{tabular} &
  \begin{tabular}[c]{@{}c@{}}0.6365\\ 0.1330\end{tabular} &
  \begin{tabular}[c]{@{}c@{}}0.6774\\ 0.1708\end{tabular} &
  \begin{tabular}[c]{@{}c@{}}0.6534\\ 0.1187\end{tabular} &
  \multicolumn{1}{c}{\begin{tabular}[c]{@{}c@{}}0.6197\\ 0.1285\end{tabular}} \\
\multicolumn{2}{c}{$n=300$} &
  \begin{tabular}[c]{@{}c@{}}Average\\ SD\end{tabular} &
  \begin{tabular}[c]{@{}c@{}}0.7865\\ 0.1141\end{tabular} &
  \begin{tabular}[c]{@{}c@{}}0.6512\\ 0.1373\end{tabular} &
  \begin{tabular}[c]{@{}c@{}}0.7525\\ 0.1839\end{tabular} &
  \begin{tabular}[c]{@{}c@{}}0.6868\\ 0.1250\end{tabular} &
  \multicolumn{1}{c}{\begin{tabular}[c]{@{}c@{}}0.6464\\ 0.1462\end{tabular}} \\
\multicolumn{2}{c}{$n=500$} &
  \begin{tabular}[c]{@{}c@{}}Average\\ SD\end{tabular} &
  \begin{tabular}[c]{@{}c@{}}0.8383\\ 0.0962\end{tabular} &
  \begin{tabular}[c]{@{}c@{}}0.7018\\ 0.1443\end{tabular} &
  \begin{tabular}[c]{@{}c@{}}0.8465\\ 0.1650\end{tabular} &
  \begin{tabular}[c]{@{}c@{}}0.7198\\ 0.1372\end{tabular} &
  \multicolumn{1}{c}{\begin{tabular}[c]{@{}c@{}}0.7260\\ 0.1602\end{tabular}} \\ \hline
\end{tabular}
\end{table}

\begin{table}[]
\caption{
Averages and standard deviations of the trace correlation $R$ for the LMOPG method for model {\bf [A1]}, {\bf [B1]}, {\bf [B2]} and {\bf [B3]} with multivariate Exponential distributed predicted variables based on 100 simulations with different sample sizes.}
\label{Exp}
        \setlength{\tabcolsep}{5.5pt} 
        \renewcommand{\arraystretch}{1.1}
        \hspace*{0.6em}
\begin{tabular}{cccccccc}
\hline
\multicolumn{1}{c}{\multirow{2}{*}{~ ~Exp~ ~}} &
  \multicolumn{2}{c}{Model} &
  \multicolumn{5}{c}{A1} \\
\multicolumn{1}{c}{} &
  \multicolumn{2}{c}{Method} &
  LMOPG &
  meanOPG &
  csOPG &
  meanMAVE &
  csMAVE \\ \hline
\multicolumn{2}{c}{$n=200$} &
  \begin{tabular}[c]{@{}c@{}}Average\\ SD\end{tabular} &
  \begin{tabular}[c]{@{}c@{}}0.8608\\ 0.0986\end{tabular} &
  \begin{tabular}[c]{@{}c@{}}0.6092\\ 0.1070\end{tabular} &
  \begin{tabular}[c]{@{}c@{}}0.7844\\ 0.1373\end{tabular} &
  \begin{tabular}[c]{@{}c@{}}0.6362\\ 0.1069\end{tabular} &
  \begin{tabular}[c]{@{}l@{}}0.8588\\ 0.1082\end{tabular} \\
\multicolumn{2}{c}{$n=300$} &
  \begin{tabular}[c]{@{}c@{}}Average\\ SD\end{tabular} &
  \begin{tabular}[c]{@{}c@{}}0.9058\\ 0.0600\end{tabular} &
  \begin{tabular}[c]{@{}c@{}}0.6019\\ 0.0876\end{tabular} &
  \begin{tabular}[c]{@{}c@{}}0.8584\\ 0.0988\end{tabular} &
  \begin{tabular}[c]{@{}c@{}}0.6208\\ 0.0905\end{tabular} &
  \begin{tabular}[c]{@{}l@{}}0.9167\\ 0.0841\end{tabular} \\
\multicolumn{2}{c}{$n=500$} &
  \begin{tabular}[c]{@{}c@{}}Average\\ SD\end{tabular} &
  \begin{tabular}[c]{@{}c@{}}0.9477\\ 0.0304\end{tabular} &
  \begin{tabular}[c]{@{}c@{}}0.6024\\ 0.1018\end{tabular} &
  \begin{tabular}[c]{@{}c@{}}0.8983\\ 0.0890\end{tabular} &
  \begin{tabular}[c]{@{}c@{}}0.6099\\ 0.0898\end{tabular} &
  \begin{tabular}[c]{@{}l@{}}0.9760\\ 0.0174\end{tabular} \\ \hline
 &
  \multicolumn{2}{c}{Model} &
  \multicolumn{5}{c}{B1} \\
\multicolumn{2}{c}{$n=200$} &
  \begin{tabular}[c]{@{}c@{}}Average\\ SD\end{tabular} &
  \begin{tabular}[c]{@{}c@{}}0.7699\\ 0.1466\end{tabular} &
  \begin{tabular}[c]{@{}c@{}}0.6604\\ 0.1476\end{tabular} &
  \begin{tabular}[c]{@{}c@{}}0.6206\\ 0.1364\end{tabular} &
  \begin{tabular}[c]{@{}c@{}}0.6556\\ 0.1366\end{tabular} &
  \begin{tabular}[c]{@{}l@{}}0.5934\\ 0.1132\end{tabular} \\
\multicolumn{2}{c}{$n=300$} &
  \begin{tabular}[c]{@{}c@{}}Average\\ SD\end{tabular} &
  \begin{tabular}[c]{@{}c@{}}0.8423\\ 0.0970\end{tabular} &
  \begin{tabular}[c]{@{}c@{}}0.7075\\ 0.1641\end{tabular} &
  \begin{tabular}[c]{@{}c@{}}0.6544\\ 0.1582\end{tabular} &
  \begin{tabular}[c]{@{}c@{}}0.7113\\ 0.1555\end{tabular} &
  \begin{tabular}[c]{@{}l@{}}0.5986\\ 0.1154\end{tabular} \\
\multicolumn{2}{c}{$n=500$} &
  \begin{tabular}[c]{@{}c@{}}Average\\ SD\end{tabular} &
  \begin{tabular}[c]{@{}c@{}}0.8959\\ 0.0784\end{tabular} &
  \begin{tabular}[c]{@{}c@{}}0.7744\\ 0.1575\end{tabular} &
  \begin{tabular}[c]{@{}c@{}}0.7233\\ 0.1844\end{tabular} &
  \begin{tabular}[c]{@{}c@{}}0.7871\\ 0.1506\end{tabular} &
  \begin{tabular}[c]{@{}l@{}}0.6609\\ 0.1480\end{tabular} \\ \hline
   &
  \multicolumn{2}{c}{Model} &
  \multicolumn{5}{c}{B2} \\
\multicolumn{2}{c}{$n=200$} &
  \begin{tabular}[c]{@{}c@{}}Average\\ SD\end{tabular} &
  \begin{tabular}[c]{@{}c@{}}0.6926\\ 0.1336\end{tabular} &
  \begin{tabular}[c]{@{}c@{}}0.5673\\ 0.0777\end{tabular} &
  \begin{tabular}[c]{@{}c@{}}0.6144\\ 0.1364\end{tabular} &
  \begin{tabular}[c]{@{}c@{}}0.5729\\ 0.0872\end{tabular} &
  \begin{tabular}[c]{@{}l@{}}0.5815\\ 0.0989\end{tabular} \\
\multicolumn{2}{c}{$n=300$} &
  \begin{tabular}[c]{@{}c@{}}Average\\ SD\end{tabular} &
  \begin{tabular}[c]{@{}c@{}}0.7489\\ 0.1121\end{tabular} &
  \begin{tabular}[c]{@{}c@{}}0.5886\\ 0.1089\end{tabular} &
  \begin{tabular}[c]{@{}c@{}}0.6367\\ 0.1514\end{tabular} &
  \begin{tabular}[c]{@{}c@{}}0.5809\\ 0.0916\end{tabular} &
  \begin{tabular}[c]{@{}l@{}}0.5967\\ 0.1155\end{tabular} \\
\multicolumn{2}{c}{$n=500$} &
  \begin{tabular}[c]{@{}c@{}}Average\\ SD\end{tabular} &
  \begin{tabular}[c]{@{}c@{}}0.8017\\ 0.1053\end{tabular} &
  \begin{tabular}[c]{@{}c@{}}0.5918\\ 0.1018\end{tabular} &
  \begin{tabular}[c]{@{}c@{}}0.6574\\ 0.1631\end{tabular} &
  \begin{tabular}[c]{@{}c@{}}0.5801\\ 0.0896\end{tabular} &
  \begin{tabular}[c]{@{}l@{}}0.6185\\ 0.1339\end{tabular} \\ \hline
 &
  \multicolumn{2}{c}{Model} &
  \multicolumn{5}{c}{B3} \\
\multicolumn{2}{c}{$n=200$} &
  \begin{tabular}[c]{@{}c@{}}Average\\ SD\end{tabular} &
  \begin{tabular}[c]{@{}c@{}}0.7400\\ 0.1282\end{tabular} &
  \begin{tabular}[c]{@{}c@{}}0.6066\\ 0.1305\end{tabular} &
  \begin{tabular}[c]{@{}c@{}}0.6452\\ 0.1557\end{tabular} &
  \begin{tabular}[c]{@{}c@{}}0.6626\\ 0.1572\end{tabular} &
  \multicolumn{1}{c}{\begin{tabular}[c]{@{}c@{}}0.6446\\ 0.1328\end{tabular}} \\
\multicolumn{2}{c}{$n=300$} &
  \begin{tabular}[c]{@{}c@{}}Average\\ SD\end{tabular} &
  \begin{tabular}[c]{@{}c@{}}0.8052\\ 0.1167\end{tabular} &
  \begin{tabular}[c]{@{}c@{}}0.6570\\ 0.1536\end{tabular} &
  \begin{tabular}[c]{@{}c@{}}0.7333\\ 0.1773\end{tabular} &
  \begin{tabular}[c]{@{}c@{}}0.6947\\ 0.1640\end{tabular} &
  \multicolumn{1}{c}{\begin{tabular}[c]{@{}c@{}}0.6826\\ 0.1530\end{tabular}} \\
\multicolumn{2}{c}{$n=500$} &
  \begin{tabular}[c]{@{}c@{}}Average\\ SD\end{tabular} &
  \begin{tabular}[c]{@{}c@{}}0.8801\\ 0.0930\end{tabular} &
  \begin{tabular}[c]{@{}c@{}}0.6814\\ 0.6507\end{tabular} &
  \begin{tabular}[c]{@{}c@{}}0.8315\\ 0.1631\end{tabular} &
  \begin{tabular}[c]{@{}c@{}}0.7869\\ 0.1757\end{tabular} &
  \multicolumn{1}{c}{\begin{tabular}[c]{@{}c@{}}0.8124\\ 0.1537\end{tabular}} \\ \hline
\end{tabular}
\end{table}

\begin{table}[]
\caption{
Averages and standard deviations of the trace correlation $R$ for the LMOPG method for model {\bf [A1]}, {\bf [B1]}, {\bf [B2]} and {\bf [B3]} with multivariate F distributed predicted variables based on 100 simulations with different sample sizes.}
\label{F}
        \setlength{\tabcolsep}{5.5pt} 
        \renewcommand{\arraystretch}{1.1}
        \hspace*{0.6em}
\begin{tabular}{cccccccc}
\hline
\multicolumn{1}{c}{\multirow{2}{*}{~ ~ ~F ~ ~ ~}} &
  \multicolumn{2}{c}{Model} &
  \multicolumn{5}{c}{A1} \\
\multicolumn{1}{c}{} &
  \multicolumn{2}{c}{Method} &
  LMOPG &
  meanOPG &
  csOPG &
  meanMAVE &
  csMAVE \\ \hline
\multicolumn{2}{c}{$n=200$} &
  \begin{tabular}[c]{@{}c@{}}Average\\ SD\end{tabular} &
  \begin{tabular}[c]{@{}c@{}}0.8400\\ 0.1250\end{tabular} &
  \begin{tabular}[c]{@{}c@{}}0.5992\\ 0.1044\end{tabular} &
  \begin{tabular}[c]{@{}c@{}}0.7912\\ 0.1365\end{tabular} &
  \begin{tabular}[c]{@{}c@{}}0.6229\\ 0.1052\end{tabular} &
  \begin{tabular}[c]{@{}l@{}}0.8605\\ 0.1144\end{tabular} \\
\multicolumn{2}{c}{$n=300$} &
  \begin{tabular}[c]{@{}c@{}}Average\\ SD\end{tabular} &
  \begin{tabular}[c]{@{}c@{}}0.8837\\ 0.1060\end{tabular} &
  \begin{tabular}[c]{@{}c@{}}0.5962\\ 0.0997\end{tabular} &
  \begin{tabular}[c]{@{}c@{}}0.8556\\ 0.1252\end{tabular} &
  \begin{tabular}[c]{@{}c@{}}0.6423\\ 0.0953\end{tabular} &
  \begin{tabular}[c]{@{}l@{}}0.9276\\ 0.0704\end{tabular} \\
\multicolumn{2}{c}{$n=500$} &
  \begin{tabular}[c]{@{}c@{}}Average\\ SD\end{tabular} &
  \begin{tabular}[c]{@{}c@{}}0.9199\\ 0.0752\end{tabular} &
  \begin{tabular}[c]{@{}c@{}}0.5867\\ 0.0953\end{tabular} &
  \begin{tabular}[c]{@{}c@{}}0.8862\\ 0.1162\end{tabular} &
  \begin{tabular}[c]{@{}c@{}}0.6288\\ 0.0926\end{tabular} &
  \begin{tabular}[c]{@{}l@{}}0.9763\\ 0.0201\end{tabular} \\ \hline
 &
  \multicolumn{2}{c}{Model} &
  \multicolumn{5}{c}{B1} \\
\multicolumn{2}{c}{$n=200$} &
  \begin{tabular}[c]{@{}c@{}}Average\\ SD\end{tabular} &
  \begin{tabular}[c]{@{}c@{}}0.7845\\ 0.1399\end{tabular} &
  \begin{tabular}[c]{@{}c@{}}0.6287\\ 0.1466\end{tabular} &
  \begin{tabular}[c]{@{}c@{}}0.5965\\ 0.1261\end{tabular} &
  \begin{tabular}[c]{@{}c@{}}0.6741\\ 0.1446\end{tabular} &
  \begin{tabular}[c]{@{}l@{}}0.5778\\ 0.1080\end{tabular} \\
\multicolumn{2}{c}{$n=300$} &
  \begin{tabular}[c]{@{}c@{}}Average\\ SD\end{tabular} &
  \begin{tabular}[c]{@{}c@{}}0.8682\\ 0.1182\end{tabular} &
  \begin{tabular}[c]{@{}c@{}}0.7056\\ 0.1675\end{tabular} &
  \begin{tabular}[c]{@{}c@{}}0.6391\\ 0.1436\end{tabular} &
  \begin{tabular}[c]{@{}c@{}}0.7112\\ 0.1641\end{tabular} &
  \begin{tabular}[c]{@{}l@{}}0.6005\\ 0.1093\end{tabular} \\
\multicolumn{2}{c}{$n=500$} &
  \begin{tabular}[c]{@{}c@{}}Average\\ SD\end{tabular} &
  \begin{tabular}[c]{@{}c@{}}0.9247\\ 0.0649\end{tabular} &
  \begin{tabular}[c]{@{}c@{}}0.7717\\ 0.1694\end{tabular} &
  \begin{tabular}[c]{@{}c@{}}0.6953\\ 0.1764\end{tabular} &
  \begin{tabular}[c]{@{}c@{}}0.7987\\ 0.1375\end{tabular} &
  \begin{tabular}[c]{@{}l@{}}0.6256\\ 0.1315\end{tabular} \\ \hline
  &
  \multicolumn{2}{c}{Model} &
  \multicolumn{5}{c}{B2} \\
\multicolumn{2}{c}{$n=200$} &
  \begin{tabular}[c]{@{}c@{}}Average\\ SD\end{tabular} &
  \begin{tabular}[c]{@{}c@{}}0.7405\\ 0.1299\end{tabular} &
  \begin{tabular}[c]{@{}c@{}}0.5941\\ 0.1113\end{tabular} &
  \begin{tabular}[c]{@{}c@{}}0.5767\\ 0.1061\end{tabular} &
  \begin{tabular}[c]{@{}c@{}}0.6047\\ 0.1058\end{tabular} &
  \begin{tabular}[c]{@{}l@{}}0.5780\\ 0.1051\end{tabular} \\
\multicolumn{2}{c}{$n=300$} &
  \begin{tabular}[c]{@{}c@{}}Average\\ SD\end{tabular} &
  \begin{tabular}[c]{@{}c@{}}0.8369\\ 0.1109\end{tabular} &
  \begin{tabular}[c]{@{}c@{}}0.6194\\ 0.1303\end{tabular} &
  \begin{tabular}[c]{@{}c@{}}0.6203\\ 0.1293\end{tabular} &
  \begin{tabular}[c]{@{}c@{}}0.6021\\ 0.1008\end{tabular} &
  \begin{tabular}[c]{@{}l@{}}0.5909\\ 0.1019\end{tabular} \\
\multicolumn{2}{c}{$n=500$} &
  \begin{tabular}[c]{@{}c@{}}Average\\ SD\end{tabular} &
  \begin{tabular}[c]{@{}c@{}}0.8848\\ 0.0842\end{tabular} &
  \begin{tabular}[c]{@{}c@{}}0.6326\\ 0.1364\end{tabular} &
  \begin{tabular}[c]{@{}c@{}}0.6403\\ 0.1459\end{tabular} &
  \begin{tabular}[c]{@{}c@{}}0.6304\\ 0.1004\end{tabular} &
  \begin{tabular}[c]{@{}l@{}}0.5901\\ 0.1019\end{tabular} \\ \hline
 &
  \multicolumn{2}{c}{Model} &
  \multicolumn{5}{c}{B3} \\
\multicolumn{2}{c}{$n=200$} &
  \begin{tabular}[c]{@{}c@{}}Average\\ SD\end{tabular} &
  \begin{tabular}[c]{@{}c@{}}0.7173\\ 0.1559\end{tabular} &
  \begin{tabular}[c]{@{}c@{}}0.5652\\ 0.0942\end{tabular} &
  \begin{tabular}[c]{@{}c@{}}0.5896\\ 0.1333\end{tabular} &
  \begin{tabular}[c]{@{}c@{}}0.6045\\ 0.1254\end{tabular} &
  \multicolumn{1}{c}{\begin{tabular}[c]{@{}c@{}}0.5798\\ 0.1112\end{tabular}} \\
\multicolumn{2}{c}{$n=300$} &
  \begin{tabular}[c]{@{}c@{}}Average\\ SD\end{tabular} &
  \begin{tabular}[c]{@{}c@{}}0.7849\\ 0.1430\end{tabular} &
  \begin{tabular}[c]{@{}c@{}}0.5900\\ 0.1221\end{tabular} &
  \begin{tabular}[c]{@{}c@{}}0.6491\\ 0.1472\end{tabular} &
  \begin{tabular}[c]{@{}c@{}}0.6389\\ 0.1331\end{tabular} &
  \multicolumn{1}{c}{\begin{tabular}[c]{@{}c@{}}0.6324\\ 0.1366\end{tabular}} \\
\multicolumn{2}{c}{$n=500$} &
  \begin{tabular}[c]{@{}c@{}}Average\\ SD\end{tabular} &
  \begin{tabular}[c]{@{}c@{}}0.8670\\ 0.1158\end{tabular} &
  \begin{tabular}[c]{@{}c@{}}0.5937\\ 0.1306\end{tabular} &
  \begin{tabular}[c]{@{}c@{}}0.7303\\ 0.1631\end{tabular} &
  \begin{tabular}[c]{@{}c@{}}0.6579\\ 0.1495\end{tabular} &
  \multicolumn{1}{c}{\begin{tabular}[c]{@{}c@{}}0.7235\\ 0.1421\end{tabular}} \\ \hline
\end{tabular}
\end{table}

\begin{table}[]
\caption{
Averages and standard deviations of the trace correlation $R$ for the LMOPG method for model {\bf [A1]}, {\bf [B1]}, {\bf [B2]} and {\bf [B3]} with multivariate Gamma distributed predicted variables based on 100 simulations with different sample sizes.}
\label{Gamma}
        \setlength{\tabcolsep}{5.5pt} 
        \renewcommand{\arraystretch}{1.1}
        \hspace*{0.6em}
\begin{tabular}{cccccccc}
\hline
\multicolumn{1}{c}{\multirow{2}{*}{Gamma}} &
  \multicolumn{2}{c}{Model} &
  \multicolumn{5}{c}{A1} \\
\multicolumn{1}{c}{} &
  \multicolumn{2}{c}{Method} &
  LMOPG &
  meanOPG &
  csOPG &
  meanMAVE &
  csMAVE \\ \hline
\multicolumn{2}{c}{$n=200$} &
  \begin{tabular}[c]{@{}c@{}}Average\\ SD\end{tabular} &
  \begin{tabular}[c]{@{}c@{}}0.7807\\ 0.1193\end{tabular} &
  \begin{tabular}[c]{@{}c@{}}0.5767\\ 0.0850\end{tabular} &
  \begin{tabular}[c]{@{}c@{}}0.7427\\ 0.1360\end{tabular} &
  \begin{tabular}[c]{@{}c@{}}0.5903\\ 0.1009\end{tabular} &
  \begin{tabular}[c]{@{}l@{}}0.7703\\ 0.1500\end{tabular} \\
\multicolumn{2}{c}{$n=300$} &
  \begin{tabular}[c]{@{}c@{}}Average\\ SD\end{tabular} &
  \begin{tabular}[c]{@{}c@{}}0.8367\\ 0.0875\end{tabular} &
  \begin{tabular}[c]{@{}c@{}}0.5792\\ 0.0852\end{tabular} &
  \begin{tabular}[c]{@{}c@{}}0.8082\\ 0.1217\end{tabular} &
  \begin{tabular}[c]{@{}c@{}}0.5803\\ 0.0830\end{tabular} &
  \begin{tabular}[c]{@{}l@{}}0.8565\\ 0.1239\end{tabular} \\
\multicolumn{2}{c}{$n=500$} &
  \begin{tabular}[c]{@{}c@{}}Average\\ SD\end{tabular} &
  \begin{tabular}[c]{@{}c@{}}0.8929\\ 0.0782\end{tabular} &
  \begin{tabular}[c]{@{}c@{}}0.5810\\ 0.0836\end{tabular} &
  \begin{tabular}[c]{@{}c@{}}0.8937\\ 0.0948\end{tabular} &
  \begin{tabular}[c]{@{}c@{}}0.6028\\ 0.0876\end{tabular} &
  \begin{tabular}[c]{@{}l@{}}0.9312\\ 0.0817\end{tabular} \\ \hline
 &
  \multicolumn{2}{c}{Model} &
  \multicolumn{5}{c}{B1} \\
\multicolumn{2}{c}{$n=200$} &
  \begin{tabular}[c]{@{}c@{}}Average\\ SD\end{tabular} &
  \begin{tabular}[c]{@{}c@{}}0.8124\\ 0.1114\end{tabular} &
  \begin{tabular}[c]{@{}c@{}}0.7434\\ 0.1475\end{tabular} &
  \begin{tabular}[c]{@{}c@{}}0.6162\\ 0.1355\end{tabular} &
  \begin{tabular}[c]{@{}c@{}}0.7395\\ 0.1496\end{tabular} &
  \begin{tabular}[c]{@{}l@{}}0.5987\\ 0.1023\end{tabular} \\
\multicolumn{2}{c}{$n=300$} &
  \begin{tabular}[c]{@{}c@{}}Average\\ SD\end{tabular} &
  \begin{tabular}[c]{@{}c@{}}0.8791\\ 0.0803\end{tabular} &
  \begin{tabular}[c]{@{}c@{}}0.8036\\ 0.1511\end{tabular} &
  \begin{tabular}[c]{@{}c@{}}0.6705\\ 0.1573\end{tabular} &
  \begin{tabular}[c]{@{}c@{}}0.8113\\ 0.1390\end{tabular} &
  \begin{tabular}[c]{@{}l@{}}0.6316\\ 0.1291\end{tabular} \\
\multicolumn{2}{c}{$n=500$} &
  \begin{tabular}[c]{@{}c@{}}Average\\ SD\end{tabular} &
  \begin{tabular}[c]{@{}c@{}}0.9302\\ 0.0373\end{tabular} &
  \begin{tabular}[c]{@{}c@{}}0.8784\\ 0.1238\end{tabular} &
  \begin{tabular}[c]{@{}c@{}}0.7574\\ 0.1825\end{tabular} &
  \begin{tabular}[c]{@{}c@{}}0.8808\\ 0.1070\end{tabular} &
  \begin{tabular}[c]{@{}l@{}}0.7016\\ 0.1582\end{tabular} \\ \hline
  &
  \multicolumn{2}{c}{Model} &
  \multicolumn{5}{c}{B2} \\
\multicolumn{2}{c}{$n=200$} &
  \begin{tabular}[c]{@{}c@{}}Average\\ SD\end{tabular} &
  \begin{tabular}[c]{@{}c@{}}0.6404\\ 0.1319\end{tabular} &
  \begin{tabular}[c]{@{}c@{}}0.5438\\ 0.0804\end{tabular} &
  \begin{tabular}[c]{@{}c@{}}0.5948\\ 0.1234\end{tabular} &
  \begin{tabular}[c]{@{}c@{}}0.5618\\ 0.0957\end{tabular} &
  \begin{tabular}[c]{@{}l@{}}0.5876\\ 0.1119\end{tabular} \\
\multicolumn{2}{c}{$n=300$} &
  \begin{tabular}[c]{@{}c@{}}Average\\ SD\end{tabular} &
  \begin{tabular}[c]{@{}c@{}}0.6863\\ 0.1323\end{tabular} &
  \begin{tabular}[c]{@{}c@{}}0.5489\\ 0.0827\end{tabular} &
  \begin{tabular}[c]{@{}c@{}}0.6077\\ 0.1252\end{tabular} &
  \begin{tabular}[c]{@{}c@{}}0.5625\\ 0.0954\end{tabular} &
  \begin{tabular}[c]{@{}l@{}}0.6078\\ 0.1210\end{tabular} \\
\multicolumn{2}{c}{$n=500$} &
  \begin{tabular}[c]{@{}c@{}}Average\\ SD\end{tabular} &
  \begin{tabular}[c]{@{}c@{}}0.7730\\ 0.1297\end{tabular} &
  \begin{tabular}[c]{@{}c@{}}0.5723\\ 0.1046\end{tabular} &
  \begin{tabular}[c]{@{}c@{}}0.6499\\ 0.1368\end{tabular} &
  \begin{tabular}[c]{@{}c@{}}0.5851\\ 0.1066\end{tabular} &
  \begin{tabular}[c]{@{}l@{}}0.6449\\ 0.1308\end{tabular} \\ \hline
 &
  \multicolumn{2}{c}{Model} &
  \multicolumn{5}{c}{B3} \\
\multicolumn{2}{c}{$n=200$} &
  \begin{tabular}[c]{@{}c@{}}Average\\ SD\end{tabular} &
  \begin{tabular}[c]{@{}c@{}}0.7678\\ 0.1114\end{tabular} &
  \begin{tabular}[c]{@{}c@{}}0.6290\\ 0.1266\end{tabular} &
  \begin{tabular}[c]{@{}c@{}}0.5915\\ 0.1043\end{tabular} &
  \begin{tabular}[c]{@{}c@{}}0.6336\\ 0.1191\end{tabular} &
  \multicolumn{1}{c}{\begin{tabular}[c]{@{}c@{}}0.5725\\ 0.0805\end{tabular}} \\
\multicolumn{2}{c}{$n=300$} &
  \begin{tabular}[c]{@{}c@{}}Average\\ SD\end{tabular} &
  \begin{tabular}[c]{@{}c@{}}0.8381\\ 0.0828\end{tabular} &
  \begin{tabular}[c]{@{}c@{}}0.6907\\ 0.1338\end{tabular} &
  \begin{tabular}[c]{@{}c@{}}0.6221\\ 0.1230\end{tabular} &
  \begin{tabular}[c]{@{}c@{}}0.6728\\ 0.1224\end{tabular} &
  \multicolumn{1}{c}{\begin{tabular}[c]{@{}c@{}}0.6068\\ 0.0991\end{tabular}} \\
\multicolumn{2}{c}{$n=500$} &
  \begin{tabular}[c]{@{}c@{}}Average\\ SD\end{tabular} &
  \begin{tabular}[c]{@{}c@{}}0.8702\\ 0.0603\end{tabular} &
  \begin{tabular}[c]{@{}c@{}}0.7647\\ 0.1353\end{tabular} &
  \begin{tabular}[c]{@{}c@{}}0.6352\\ 0.1330\end{tabular} &
  \begin{tabular}[c]{@{}c@{}}0.7261\\ 0.1318\end{tabular} &
  \multicolumn{1}{c}{\begin{tabular}[c]{@{}c@{}}0.6056\\ 0.1065\end{tabular}} \\ \hline
\end{tabular}
\end{table}

From Table \ref{Chisq}-\ref{Gamma}, once again, the performance of the methods considered here improved with the increase in sample size. The LMOPG method performs well in almost all the settings considered here, with skewed residuals distribution.  The csOPG or csMAVE performs better than meanOPG or meanMAVE under model {\bf[A1]} and it is also consistent with the theoretical results. The five methods perform better under model {\bf[B1]} than  model {\bf[B2]}. Although the absolute value function is not differentiable, the predictor variables take positive values under all four distributions considered. Hence, model {\bf[B1]} is differentiable and the model {\bf[B2]} is not differentiable due to the residuals. Since we suppose the link function is differentiable, the better performances obtained under model {\bf[B1]} are obvious.

In particular, from table \ref{Chisq}, compared to the csMAVE method, the LMOPG underperforms under model {\bf[A1]}, {\bf[B1]} and {\bf[B2]} with four distributions. This may be due to the fact that csMAVE directly optimizes to estimate the basis matrix $\text{B}$. It is may be more accurate compared to to estimate the derivatives and then make the eigendecomposition. The simulation studies in \citep{xia2002adaptive} and \citep{opg2007} have the similar results. And the LMOPG underperforms under model {\bf[A1]} with the multivariate Chi-squares distribution at 100 sample size, but  outperforms at 500 sample size.

 In summary, as demonstrated by the simulation results in Sections \ref{sec4.1}--\ref{sec4.2}, the proposed LMOPG method exhibits promising performance when the distribution of residuals is not symmetric.

\section{Real data analysis}
The Gas Turbine CO and NOx Emission dataset presented in \citep{misc_gas_turbine_co_and_nox_emission_data_set_551} is a publicly available dataset that contains five sub-dataset by year for five years, 2011 to 2015.  The dataset comes from the same power plant as the dataset which come from a gas turbine located in Turkey's north western region for the purpose of studying flue gas emissions, namely CO and NOx (NO + NO2).
The variables in this dataset are: Ambient temperature (AT); Ambient pressure (AP); Ambient humidity (AH); Air filter difference pressure (AFDP); Gas turbine exhaust pressure (GTEP); Turbine inlet temperature (TIT); Turbine after temperature (TAT); Compressor discharge pressure (CDP); Turbine energy yield (TEY); Carbon monoxide (CO) and Nitrogen oxides (NOx). The variables TEY, CO and NOx can be considered as response variables and the remaining 8 variables can be considered as predictor variables. As described for the dataset \citep{misc_gas_turbine_co_and_nox_emission_data_set_551}, the dataset can be well used for predicting turbine energy yield (TEY) using ambient variables as features. Hence, we only consider TEY as the response variable. Furthermore, We think of this problem as a regression problem and apply the proposed LMOPG method to botain the coefficient of regression, with $1,500$ observations in the sub-dataset of 2011. And the first $1,000$ observations are used for training and the next $500$ for testing.

\begin{figure}[ht]
\centering
\includegraphics[width=0.7\textwidth]{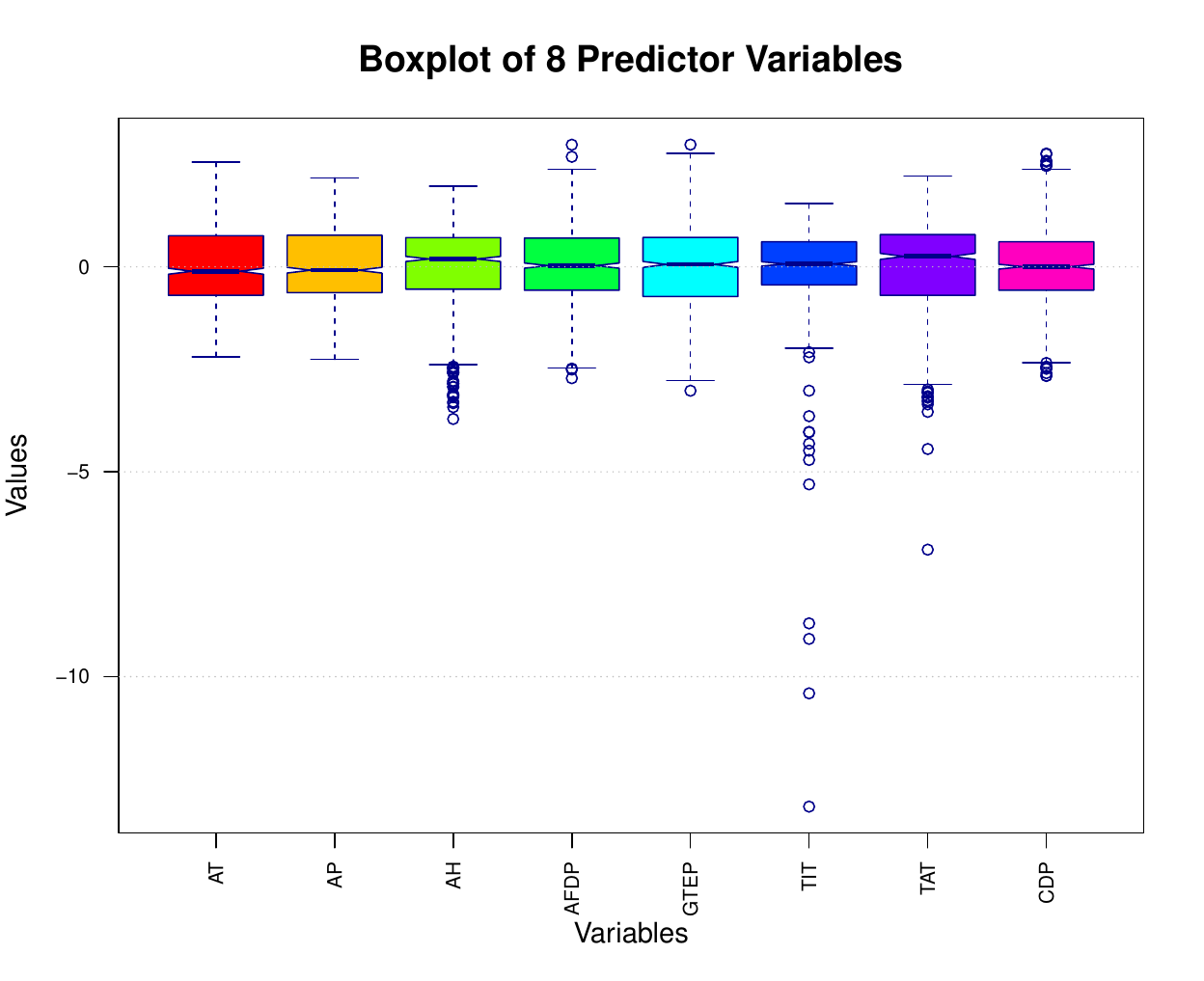}
    \caption{The Box plot of $8$ predictor variables.}
\label{fig:Boxplot}
\end{figure}

\begin{figure}ht]
\centering
\includegraphics[width=0.7\textwidth]{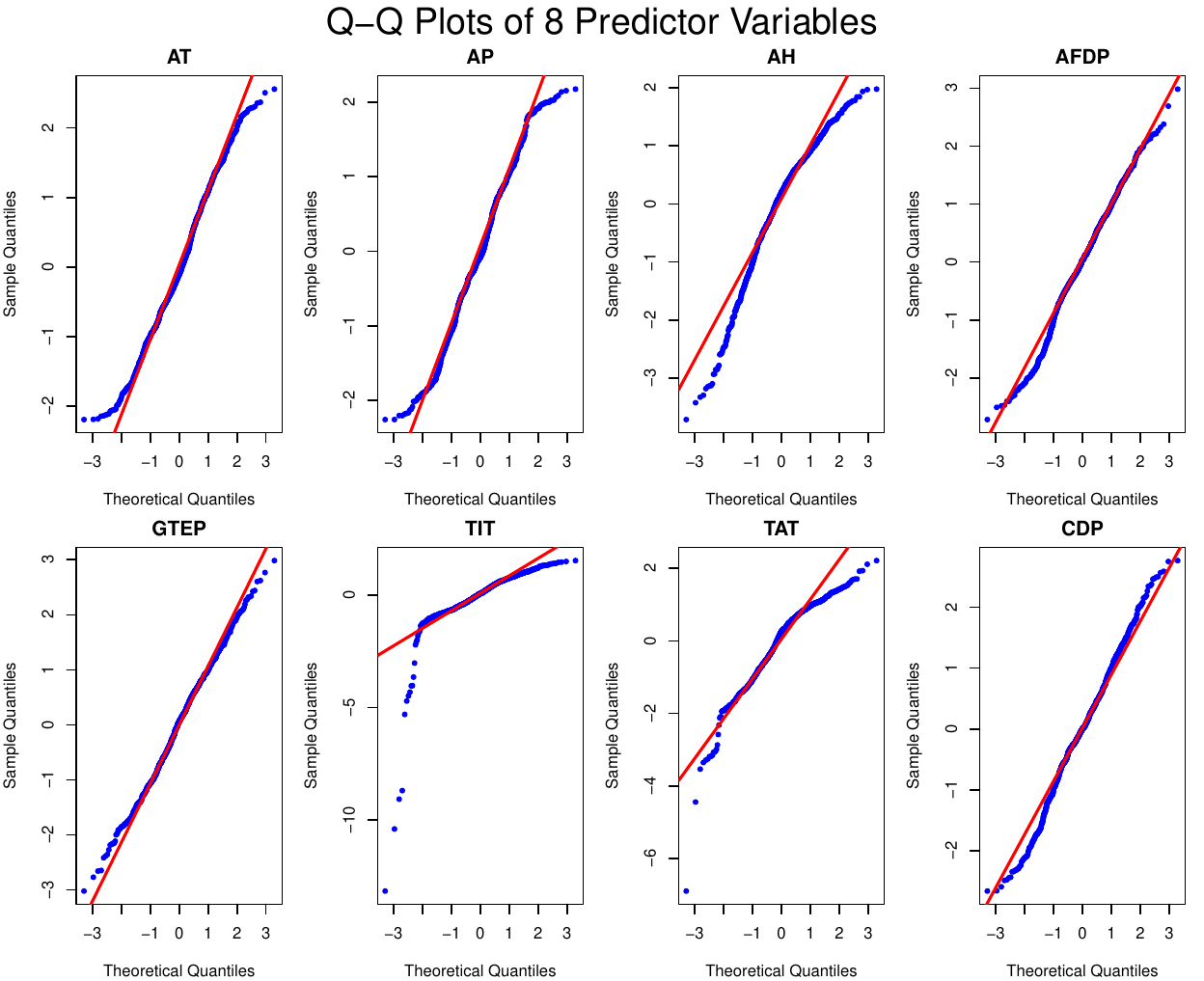}
    \caption{The Q-Q plot of $8$ predictor variables.}
\label{fig:qqplot}
\end{figure}

Next, we take a simple analysis of the dataset. First, to assess the symmetry of the distributions of these 8 predictor variables in the sub-dataset of 2011, we provide the comparative boxplots in Figure \ref{fig:Boxplot}, after standardizing the data. We can find that AH, TIT and TAT are clearly asymmetric. Then, we assess the normality of the $8$ predictor variables using the hypothesis testing approach and graphical approach, namely normal quantile-quantile (Q-Q) plot and the Shapiro-Wilk test. The normal Q-Q plots are depicted in Figure \ref{fig:qqplot}. Table \ref{SWnormal} presents the Shapiro-Wilk statistics for predictor variables. From the results of Q-Q plot and Shapiro-Wilk tests, almost all predictor variables do not follow normal distribution.

\begin{figure}[ht]
\centering
\includegraphics[width=0.7\textwidth]{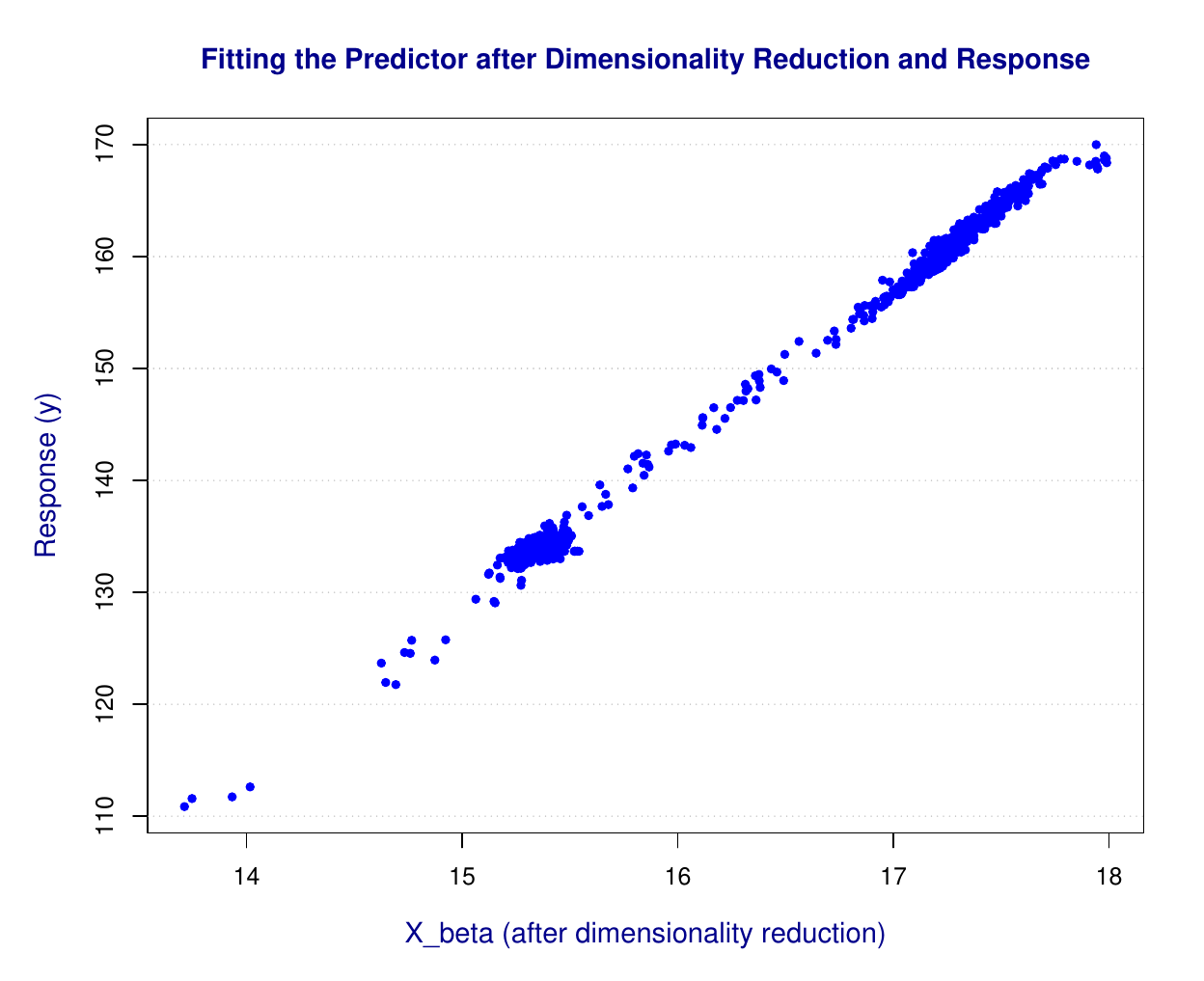}
    \caption{Fitting the Predictor with reducing dimensionality and Response.}
\label{fig:fitting}
\end{figure}

Then, we apply the LMOPG method to address the regression problem at hand. The first step involves determining the dimension of the central subspace. We consider evaluating the proportion that an eigenvalue accounts for all the eigenvalues to determine the dimension of the central space, which is similar to the way of determining the number of principal components in PCA. The results show that the proportion of the first eigenvalues is approximately equal to 1, which indicate that the dimension of central space can be considered as $1$.  The second step is to estimate the basis $\hat{\beta}_{1}$ of the central space. The result of $\hat{\beta}_{1}$ is $(0.0072, -0.0055, -0.0474,  0.0286,  0.3115,  0.5647, -0.7574,  0.0849)$. In the third step we draw a scatter plot of the relationship between the predictor variables after reducing dimension ($\hat{\beta}_{1}^\top\bm{X}$) and response variables (TEY, $\textsl{y}$). The scatter plot shows in Figure \ref{fig:fitting}. We can see that it is like a linear relationship. So, the fourth step is to perform a linear regression. The coefficients of $\hat{\beta}_{1}^\top\bm{X}$ is $14.0690$ and the intercept is $-82.0064$. All their $p$-values are less than $0.001$. The adjust $R$-squared is $0.9965$ and the $F$-statistic is $286,100$, the corresponding $p$-value is less than $0.001$. The above results indicate that the regression coefficients are significant. The final step is to make predictions and compare them with the test set. The results in test set are as follows: the adjust $R$-squared is $0.9959$, the mean squared error is $0.7922$, the root mean squared error is $0.8901$ and the $F$-statistic is $70.2857$, the corresponding $p$-value is less than $0.001$. The results show that our predictions turned out to be good.

\begin{table}[ht]
\centering
  \caption{The test statistics and $p$-values of Shapiro-Wilk normality tests for the 8 variables.}
  \label{SWnormal}
  \setlength{\tabcolsep}{4pt} 
  \renewcommand{\arraystretch}{1.5}
  {\small
\begin{tabular}{ccccccccc}
\hline
Predictor Variables & AT       & AP       & AH       & AFDP     & GTEP       & TIT      & TAT      & CDP         \\ \hline
Test Stat.          & 0.9896   & 0.9873   & 0.9548   & 0.9922   & 0.9963     & 0.6670   & 0.9409   & 0.9949      \\
$p$-value           & {\footnotesize <0.001} & {\footnotesize <0.001} & {\footnotesize <0.001} & {\footnotesize <0.001} & {\footnotesize <0.05} & {\footnotesize <0.001} & {\footnotesize <0.001} & {\footnotesize <0.005} \\ \hline
\end{tabular}}
\end{table}

\section{Conclusion}
In this paper, we propose a local modal out product gradient (LMOPG) method, a novel approach for dimension reduction tailored for high-dimensional data. Drawing inspiration from the modal linear regression (MODLR) and out product gradient (OPG) method,LMOPG is devised to handle the complexities of such datasets effectively.

The core principle of LMOPG lies in modeling the plurality of conditional distributions whose first-order derivatives contain information about the basis of the central subspace. Unlike the meanOPG method, which may miss some information in the center subspace in some cases and may not be robust because of modeling the mean, LMOPG aims at capturing a comprehensive estimate of this central subspace. The LMOPG method achieves this by estimating the first order derivatives. Moreover, this paper presents a fundamental theorem affirming the efficacy of LMOPG in achieving substantial dimension reduction. Additionally, we provide a simple algorithm for implementing LMOPG. We delve into the asymptotic behavior of the LMOPG estimator, elucidating its convergence rate in high-dimensional scenarios.

Our simulation results underscore the superiority of LMOPG in enhancing estimation accuracy for different distributed data with the skewed residuals distribution. Overall, the proposed LMOPG method furnishes invaluable tools for dimension reduction and analysis of high-dimensional, exhibiting resilience even in the face of deviations from the non-differentiable link function.

\subsection*{Competing interests}
The authors declare there are no conflict of interests.

\subsection*{Funding}
The research is supported by the National Natural Science Foundation of China (No. 12101146). 

\appendix
\section{Proofs of main results}
\subsection*{Proof of Lemma \ref{Zhidaolem}}
\begin{proof}
According to \eqref{ChainRul}, we have
$$ E\left( \frac{\partial m(\bm{X})}{\partial \bm{X}} \frac{\partial m(\bm{X})}{\partial \bm{X}^\top} \right)=\text{B}_{0}E\left(\frac{\partial m(\text{B}_{0}^\top\bm{X})}{\partial (\text{B}_{0}^\top\bm{X})}\frac{\partial m(\text{B}_{0}^\top\bm{X})}{\partial (\bm{X}^\top\text{B}_{0})}\right)\text{B}_{0}^\top.$$
Because $E((\partial m(\bm{X})/\partial \bm{X})(\partial m(\bm{X})/\partial \bm{X}^\top))$ is a semi-positive defined matrix, for any non-zero vector $ \eta \in \mathbb{R}^{p}$, have
$$\eta^\top E\left( \frac{\partial m(\bm{X})}{\partial \bm{X}} \frac{\partial m(\bm{X})}{\partial \bm{X}^\top } \right)\eta=\eta^\top \text{B}_{0}E\left(\frac{\partial m(\text{B}_{0}^\top \bm{X})}{\partial (\text{B}_{0}^\top \bm{X})}\frac{\partial m(\text{B}_{0}^\top \bm{X})}{\partial (\bm{X}^\top \text{B}_{0})}\right)\text{B}_{0}^\top \eta\geq0.$$
Let $\eta^*=\text{B}_{0}^\top \eta$. Because $\text{B}_{0}$ is a full-column rank matrix, the rows of $\text{B}_{0}^\top $ are linear independent. Then $\eta^*$ can be any vector in $\mathbb{R}^{d}$. And the rank of \\ $E((\partial m(\bm{X})/\partial \bm{X})(\partial m(\bm{X})/\partial \bm{X}^\top ))$ is $d$. Hence, for any non-zero vector $\eta^*\in \mathbb{R}^{d}$, we have
$$\eta^{*^\top }E\left(\frac{\partial m(\text{B}_{0}^\top \bm{X})}{\partial (\text{B}_{0}^\top \bm{X})}\frac{\partial m(\text{B}_{0}^\top \bm{X})}{\partial (\bm{X}^\top \text{B}_{0})}\right)\eta^*>0.$$
That is, $E[(\partial m(\text{B}_{0}^\top \bm{X})/\partial \text{B}_{0}^\top \bm{X})(\partial m(\text{B}_{0}^\top \bm{X})/\partial \bm{X}^\top \text{B}_{0})]$ is a positive defined matrix. In other words, we can obtain
$$\textup{Span}\left [ E\left( \frac{\partial m(\bm{X})}{\partial \bm{X}} \frac{\partial m(\bm{X})}{\partial \bm{X}^\top } \right) \right ]=\textup{Span}\left[\text{B}_{0}E\left(\frac{\partial m(\text{B}_{0}^\top \bm{X})}{\partial (\text{B}_{0}^\top \bm{X})}\frac{\partial m(\text{B}_{0}^\top \bm{X})}{\partial (\bm{X}^\top \text{B}_{0})}\right)\text{B}_{0}^\top \right]=\textup{Span}(\textup{B}_{0}).$$

\end{proof}

\subsection*{Proof of Theorem \ref{iterTh}}
\begin{proof}
\begin{align*}
\textup{log}&\left(\textup{L}\left(\theta^{(t+1)}\right) \right)-\textup{log}\left(\textup{L}\left(\theta^{(t)}\right) \right)\\
&=\textup{log}\left[\sum_{i=1}^{n}K_{h_{1}}(\bm{X}_{i}-X_0)\phi_{h_{2}}\left(\textsl{y}_{i}-b_{0}^{(t+1)}-\bm{b}^{(t+1)^\top }(\bm{X}_{i}-X_0) \right) \right]\\
& \qquad \qquad -\textup{log}\left[\sum_{i=1}^{n}K_{h_{1}}(\bm{X}_{i}-X_0)\phi_{h_{2}}\left(\textsl{y}_{i}-b_{0}^{(t)}-\bm{b}^{(t)^\top }(\bm{X}_{i}-X_0) \right) \right]\\
&=\textup{log}\left[ \sum_{i=1}^{n}\frac{K_{h_{1}}(\bm{X}_{i}-X_0)\phi_{h_{2}}\left(\textsl{y}_{i}-b_{0}^{(t+1)}-\bm{b}^{(t+1)^\top }(\bm{X}_{i}-X_0) \right)}{\sum_{i=1}^{n}K_{h_{1}}(\bm{X}_{i}-X_0)\phi_{h_{2}}\left(\textsl{y}_{i}-b_{0}^{(t)}-\bm{b}^{(t)^\top }(\bm{X}_{i}-X_0) \right)} \right]\\
&=\textup{log}\left[ \sum_{i=1}^{n}\frac{K_{h_{1}}(\bm{X}_{i}-X_0)\phi_{h_{2}}\left(\textsl{y}_{i}-b_{0}^{(t)}-\bm{b}^{(t)^\top }(\bm{X}_{i}-X_0) \right)}{\sum_{i=1}^{n}K_{h_{1}}(\bm{X}_{i}-X_0)\phi_{h_{2}}\left(\textsl{y}_{i}-b_{0}^{(t)}-\bm{b}^{(t)^\top }(\bm{X}_{i}-X_0) \right)}\right.\\
&\qquad \qquad \qquad \left. \cdot \frac{K_{h_{1}}(\bm{X}_{i}-X_0)\phi_{h_{2}}\left(\textsl{y}_{i}-b_{0}^{(t+1)}-\bm{b}^{(t+1)^\top }(\bm{X}_{i}-X_0) \right)}{K_{h_{1}}(\bm{X}_{i}-X_0)\phi_{h_{2}}\left(\textsl{y}_{i}-b_{0}^{(t)}-\bm{b}^{(t)^\top }(\bm{X}_{i}-X_0) \right)} \right]\\
&=\textup{log} \left[\sum_{i=1}^{n}\textup{W}(\textit{i}\mid \theta^{(t)}) \frac{\phi_{h_{2}}\left(\textsl{y}_{i}-b_{0}^{(t+1)}-\bm{b}^{(t+1)^\top }(\bm{X}_{i}-X_0) \right)}{\phi_{h_{2}}\left(\textsl{y}_{i}-b_{0}^{(t)}-\bm{b}^{(t)^\top }(\bm{X}_{i}-X_0) \right)}\right].
\end{align*}
According to the Jensen's inequality, we can get
\begin{align*}
\textup{log}\left(\textup{L}\left(\theta^{(t+1)}\right) \right)&-\textup{log}\left(\textup{L}\left(\theta^{(t)}\right) \right)\\  
&\geq\sum_{i=1}^{n}\textup{W}(\textit{i}\mid \theta^{(t)})\log\left[ \frac{\phi_{h_{2}}\left(\textsl{y}_{i}-b_{0}^{(t+1)}-\bm{b}^{(t+1)^\top }(\bm{X}_{i}-X_0) \right)}{\phi_{h_{2}}\left(\textsl{y}_{i}-b_{0}^{(t)}-\bm{b}^{(t)^\top }(\bm{X}_{i}-X_0) \right)}\right].
\end{align*}
According to the property of Step {\footnotesize 9} in Algorithm \ref{LMOPG}, we have The right-hand side of the above inequality is greater than 0, that is,
$$\textup{log}\left(\textup{L}\left(\theta^{(t+1)}\right) \right)\geq \textup{log}\left(\textup{L}\left(\theta^{(t)}\right) \right).$$
And thus, 
$$\textup{L}\left(\theta^{(t+1)}\right) \geq \textup{L}\left(\theta^{(t)}\right).$$
\end{proof}

\subsection*{Proof of Theorem \ref{consistentTheta}}
\begin{proof}
Shorten $K_{h_{1}}(\bm{X}_{i}-X_0)$ to $K_{h_{1}}$. Because $\phi_{h_{2}}(t)$ is the Gaussian kernel, we have
$$\phi_{h_{2}}'(t)=-\frac{t}{h_{2}^{3}}\phi\left(\frac{t}{h_{2}}\right), \ \phi_{h_{2}}''(t)=-\frac{1}{h_{2}^{3}}\left(\frac{t^{2}}{h_{2}^{2}}-1\right)\phi\left(\frac{t}{h_{2}}\right), $$ and
$$ \phi_{h_{2}}'''(t)=-\frac{1}{h_{2}^{4}}\left(\frac{3t}{h_{2}}-\frac{t^{3}}{h_{2}^{3}}\right)\phi\left(\frac{t}{h_{2}}\right).$$
Denote $\alpha_{n}=(nh_{1}^ph_{2}^{3})^{-1/2}+h_{1}^{2}+h_{2}^{2}$. It is sufficient to show that for any given $\eta  > 0$, there exists
a large constant $c$ and a $p+1$-dimensional non-random vector $\mu$ such that
\begin{equation}
\label{promianeq}
    P\left\{\mathop{\text{sup}}_{\|\mu\|_2=c}\textup{L}(\theta^{*}+\alpha_{n}\mu)<\textup{L}(\theta^{*})\right\}\geq 1-\eta,
\end{equation}
where $\textup{L}(\cdot)$ is defined in \eqref{Obfunc}. By using Taylor expansion, it follows that
\begin{align*}
    &\textup{L}(\theta^{*}+\alpha_{n}\mu)-\textup{L}(\theta^{*})\\
    &=\frac{1}{n}\sum_{i=1}^{n}K_{h_{1}}\left[-\phi_{h_{2}}'\left(\varepsilon_{i}^{*}+R(\bm{X}_{i})\right)\alpha_{n}\mu^\top \bm{\text{X}}_{i}^{*}+\frac{1}{2}\phi_{h_{2}}''\left(\varepsilon_{i}^{*}+R(\bm{X}_{i})\right)\alpha_{n}^{2}(\mu^\top \bm{\text{X}}_{i}^{*})^{2}\right.\\
    & \qquad \qquad \qquad\left.-\frac{1}{6}\phi_{h_{2}}'''\left(\zeta _{i}\right)\alpha_{n}^{3}(\mu^\top \bm{\text{X}}_{i}^{*})^{3} \right]
     \equiv \mathbb{I}_{1}+\mathbb{I}_{2}+\mathbb{I}_{3}.
\end{align*}
where $\zeta _{i}$ is between $\varepsilon_{i}^{*}+R(\bm{X}_{i})$ and $\varepsilon_{i}^{*}+R(\bm{X}_{i})+\alpha_{n}\mu^\top \bm{\text{X}}_{i}^{*}$. Next, we will calculate $\mathbb{I}_{1}$, $\mathbb{I}_{2}$ and $\mathbb{I}_{3}$ by using the result $\bm{X}=E(\bm{X})+O_{p}([\textup{Var}(\bm{X})]^{1/2})$.
First, we calculate the mean of $\mathbb{I}_{1}$,
\begin{align*}
    E(\mathbb{I}_1)&=-\frac{1}{n}\sum_{i=1}^{n}E\left[K_{h_1}\phi_{h_{2}}'\left(\varepsilon_{i}^{*}+R(\bm{X}_{i})\right)\alpha_{n}\mu^\top \bm{\text{X}}_{i}^{*}\right]\\
    &=-\alpha_{n}\mu^\top E\left[K_{h_1}\bm{\text{X}}_{i}^{*}E\left(\phi_{h_{2}}'\left(\varepsilon_{i}^{*}+R(\bm{X}_{i})\right)\mid \bm{X}_{i}\right)\right].
\end{align*}
Next, calculate $E\left(\phi_{h_{2}}'\left(\varepsilon_{i}^{*}+R(\bm{X}_{i})\right)\mid \bm{X}_{i}\right)$, the region of integration involved below is negative infinity to positive infinity.
\begin{align*}
    E\left(\phi_{h_{2}}'\left(\varepsilon_{i}^{*}+R(\bm{X}_{i})\right)\mid \bm{X}_{i}\right)&=\int \phi_{h_{2}}'\left(\varepsilon_{i}^{*}+R(\bm{X}_{i})\right)g(\varepsilon_i^*\mid \bm{X}_i)\text{d}\varepsilon_i^*\\
    &=-\int\frac{\varepsilon_{i}^{*}+R(\bm{X}_{i})}{h_2^3}\phi\left(\frac{\varepsilon_{i}^{*}+R(\bm{X}_{i})}{h_2}\right)g(\varepsilon_i^*\mid \bm{X}_i)\text{d}\varepsilon_i^*\\
&=-\frac{1}{h_{2}}\int t \phi(t) g(th_2-R(\bm{X}_{i}))\mid \bm{X}_{i})\text{d}t.
\end{align*}
Suppose $R(\bm{X}_i)=O_p(h_2)$  and A(2) hold,  for $g(th_2-R(\bm{X}_{i}))\mid \bm{X}_{i})$, by Taylor expansion, we have
\begin{align*}
    &g(th_2-R(\bm{X}_{i}))\mid \bm{X}_{i})=g(0\mid \bm{X}_i)+h_2g'(0\mid \bm{X}_i)\left(t-\frac{R(\bm{X}_i)}{h_2}\right)\\
    &+\frac{1}{2}h_2^2g''(0\mid \bm{X}_i)\left(t-\frac{R(\bm{X}_i)}{h_2}\right)^2+\frac{1}{6}h_2^3g'''(0\mid \bm{X}_i)\left(t-\frac{R(\bm{X}_i)}{h_2}\right)^3+o\left(\left(t-\frac{R(\bm{X}_i)}{h_2}\right)^3 \right).
\end{align*}
So, 
\begin{align*}
    E\left(\phi_{h_{2}}'\left(\varepsilon_{i}^{*}+R(\bm{X}_{i})\right)\mid \bm{X}_{i}\right)=g''(0\mid \bm{X}_i)R(\bm{X}_i)-\frac{h_2^2}{2}g'''(0\mid \bm{X}_i)+o_p(1)
\end{align*}
and
\begin{align*}
E(\mathbb{I}_1)&=\alpha_{n}\mu^\top E\left[K_{h_1}\bm{\text{X}}_{i}^{*}\left(\frac{h_2^2}{2}g'''(0\mid \bm{X}_i)-g''(0\mid \bm{X}_i)R(\bm{X}_i)\right)\right]+o_p(1)\\
&=\textup{I}_1+\textup{I}_2.
\end{align*}
\begin{align*}
    \textup{I}_1&=\frac{\alpha_{n}\mu^\top h_2^2}{2h_1^p}\int \cdots \int K\left( \frac{\bm{x}_{i1}-\bm{x}_{01}}{h_1},\cdots,\frac{\bm{x}_{ip}-\bm{x}_{0p}}{h_1}\right)\bm{\text{X}}_i^* g'''(0\mid \bm{X}_i)f(\bm{X}_{i})\text{d}\bm{x}_{i1}\cdots \text{d}\bm{x}_{ip}\\
    &=\frac{h_2^2}{2}\alpha_{n}\mu^\top g'''(0\mid X_{0})f(X_0)\int \cdots \int K\left( t_1,\cdots,t_p\right) \begin{pmatrix}
1\\ 
t_1\\ 
\vdots \\ 
t_p
\end{pmatrix}\text{d}t_1\cdots \text{d}t_p+o(1)\\
    &\equiv \frac{h_2^2}{2}\alpha_{n}\mu^\top g'''(0\mid X_{0})f(X_0)\upsilon _{\textup{I}_1} +o(1).
\end{align*}
According to the definition of $R(\bm{X}_{i})$ and \eqref{TylEpa}, we have 
\begin{align*}
    \textup{I}_2&=\frac{\alpha_{n}\mu^\top }{h_1^p}\int \cdots \int K\left( \frac{\bm{x}_{i1}-\bm{x}_{01}}{h_1},\cdots,\frac{\bm{x}_{ip}-\bm{x}_{0p}}{h_1}\right) \bm{\text{X}}_i^*g''(0\mid \bm{X}_i)\\
    &\qquad \qquad \qquad \left(\frac{1}{2}(\bm{X}_i-X_{0})^\top m^{(2)}(\bm{X}_\xi )(\bm{X}_i-X_{0})\right)f(\bm{X}_{i})\text{d}\bm{x}_{i1}\cdots \text{d}\bm{x}_{ip}\\
    &=\frac{\alpha_{n}\mu^\top }{2h_1^p}\int \cdots \int K\left( \frac{\bm{x}_{i1}-\bm{x}_{01}}{h_1},\cdots,\frac{\bm{x}_{ip}-\bm{x}_{0p}}{h_1}\right) \bm{\text{X}}_i^*g''(0\mid \bm{X}_i)\\
    &\qquad \qquad \qquad \frac{(\bm{X}_i-X_{0})^\top }{h_1}h_1I_{p}m^{(2)}(\bm{X}_\xi )I_{p}h_1\frac{(\bm{X}_i-X_{0})}{h_1}f(\bm{X}_{i})\text{d}\bm{x}_{i1}\cdots \text{d}\bm{x}_{ip}\\
    &=\frac{h_1^2}{2} \alpha_{n}\mu^\top g''(0\mid X_{0})f(X_0)\\
    &\times\int \cdots \int K\left( t_1,\cdots,t_p\right) \begin{pmatrix}
1\\ 
t_1\\ 
\vdots \\ 
t_p
\end{pmatrix}(t_1,\cdots,t_p)m^{(2)}(X_{0} )\begin{pmatrix}
t_1\\ 
\vdots \\ 
t_p
\end{pmatrix}\text{d}t_1\cdots \text{d}t_p+o(1)\\
&\equiv\frac{h_1^2}{2} \alpha_{n}\mu^\top g''(0\mid X_{0})f(X_0)\upsilon _{\textup{I}_2} +o(1).
\end{align*}
So, we can obtain 
$$E(\mathbb{I}_1)=O\left(\alpha_n c\left(h_1^2+h_2^2\right)\right).$$
Next, we calculate the variance of $\mathbb{I}_{1}$,
\begin{align*}
    &\text{Var}(\mathbb{I}_1)=\frac{1}{n^2}\text{Var}\left[\sum_{i=1}^{n}\left(K_{h_1}\phi_{h_{2}}'\left(\varepsilon_{i}^{*}+R(\bm{X}_{i})\right)\alpha_{n}\mu^\top \bm{\text{X}}_{i}^{*}\right) \right]\\
    &=\frac{1}{n^2}\sum_{i=1}^{n} \text{Var}\left[K_{h_1}\phi_{h_{2}}'\left(\varepsilon_{i}^{*}+R(\bm{X}_{i})\right)\alpha_{n}\mu^\top \bm{\text{X}}_{i}^{*}\right]\\
    & \qquad + \frac{1}{n^2}\sum_{k=1}^{n}\sum_{l=1}^{n}\text{Cov}\left[K_{h_1}\phi_{h_{2}}'\left(\varepsilon_{k}^{*}+R(\bm{X}_{k})\right)\alpha_{n}\mu^\top \bm{\text{X}}_{k}^{*},K_{h_1}\phi_{h_{2}}'\left(\varepsilon_{l}^{*}+R(\bm{X}_{l})\right)\alpha_{n}\mu^\top \bm{\text{X}}_{l}^{*}\right]\\
    &=\frac{1}{n} \text{Var}\left[K_{h_1}\phi_{h_{2}}'\left(\varepsilon_{i}^{*}+R(\bm{X}_{i})\right)\alpha_{n}\mu^\top \bm{\text{X}}_{i}^{*}\right]\\
&=\frac{1}{n}E\left[K_{h_1}^2\phi_{h_{2}}'^2\left(\varepsilon_{i}^{*}+R(\bm{X}_{i})\right)\alpha_{n}^2\left(\mu^\top \bm{\text{X}}_{i}^{*}\right)^2 \right]-\frac{1}{n}E(\mathbb{I}_1)^2.
\end{align*}
By the the law of iterated expectation, we have
$$E\left[K_{h_1}^2\phi_{h_{2}}'^2\left(\varepsilon_{i}^{*}+R(\bm{X}_{i})\right)\alpha_{n}^2\left(\mu^\top \bm{\text{X}}_{i}^{*}\right)^2 \right]=E\left[K_{h_1}^2\alpha_{n}^2\left(\mu^\top \bm{\text{X}}_{i}^{*}\right)^2E\left(\phi_{h_{2}}'^2\left(\varepsilon_{i}^{*}+R(\bm{X}_{i})\right)\mid \bm{X}_{i}\right) \right].$$
By the similar way, we have
\begin{align*}
E\left(\phi_{h_{2}}'^2(\varepsilon_{i}^{*}+R(\bm{X}_{i}))\mid \bm{X}_{i}\right)&=\frac{1}{4\sqrt{\pi}h_2^3}g(0\mid \bm{X}_i)+\frac{3}{16\sqrt{\pi}h_2}g''(0\mid \bm{X}_i)\\
 &+\frac{3}{16\sqrt{\pi}h_2}g'''(0\mid \bm{X}_i)R(\bm{X}_i)+o_p(1).
\end{align*}
So, 
\begin{align*}
&E\left[K_{h_1}^2\phi_{h_{2}}'^2\left(\varepsilon_{i}^{*}+R(\bm{X}_{i})\right)\alpha_{n}^2\left(\mu^\top \bm{\text{X}}_{i}^{*}\right)^2 \right]\\
&=E\left[K_{h_1}^2\alpha_{n}^2\left(\mu^\top \bm{\text{X}}_{i}^{*}\right)^2\left(\frac{1}{4\sqrt{\pi}h_2^3}g(0\mid \bm{X}_i)+\frac{3}{16\sqrt{\pi}h_2}g''(0\mid \bm{X}_i)\right.\right.\\
&\qquad \qquad +\left. \left.\frac{3}{16\sqrt{\pi}h_2}g'''(0\mid \bm{X}_i)R(\bm{X}_i)+o_p(1)\right) \right]+o_p(1)\\
&\equiv \text{S}_{1}+\text{S}_{2}+\text{S}_{3}+o_p(1).
\end{align*}
In turn, it is easy to get the following
\begin{align*}
    \text{S}_{1}&=\frac{\alpha_{n}^2}{4\sqrt{\pi}h_2^3}\mu^\top E\left[K_{h_1}^2g(0\mid \bm{X}_i)\bm{\text{X}}_{i}^{*}\bm{\text{X}}_{i}^{*^\top }\right]\mu\\
    &=\frac{\alpha_{n}^2}{4\sqrt{\pi}h_2^3h_{1}^{2p}}\mu^\top \left[\int \cdots \int 
    K^2\left( \frac{\bm{x}_{i1}-\bm{x}_{01}}{h_1},\cdots,\frac{\bm{x}_{ip}-\bm{x}_{0p}}{h_1}\right)g(0\mid \bm{X}_i)\right.\\
    &\left.\qquad \qquad \qquad\cdot \bm{\text{X}}_{i}^{*}\bm{\text{X}}_{i}^{*^\top }f(\bm{X}_i)\text{d}\bm{x}_{i1}\cdots \text{d}\bm{x}_{ip}\right]\mu\\
    &=\frac{\alpha_{n}^2}{4\sqrt{\pi}h_2^3h_{1}^{p}}\mu^\top g(0\mid X_{0})f(X_{0})\\
    &\qquad \qquad \qquad\cdot\int \cdots \int K^2\left( t_1,\cdots,t_p\right) \begin{pmatrix}
1\\ 
t_1\\ 
\vdots \\ 
t_p
\end{pmatrix}(1,t_1,\dots,t_p)\text{d}t_1\cdots \text{d}t_p\cdot\mu+o(1)\\
&\equiv \frac{\alpha_{n}^2}{4\sqrt{\pi}h_2^3h_{1}^{p}}g(0\mid X_{0})f(X_{0})\mu^\top \Delta_1\mu+o(1),
\end{align*}
\begin{align*}
     \text{S}_{2}=\frac{3\alpha_{n}^2}{16\sqrt{\pi}h_2h_{1}^{p}}g''(0\mid X_{0})f(X_{0})\mu^\top \Delta_1\mu+o(1),
\end{align*}
and
\begin{align*}
&\text{S}_{3}=\frac{3\alpha_{n}^2\mu^\top }{32\sqrt{\pi}h_2h_{1}^{p}}g'''(0\mid X_{0})f(X_{0})
\int \cdots \int K^2\left( t_1,\cdots,t_p\right) \begin{pmatrix}
1\\ 
t_1\\ 
\vdots \\ 
t_p
\end{pmatrix}(t_1,\cdots,t_p)\\
&\qquad \qquad \qquad \cdot h_1I_pm^{(2)}(X_{0} )I_ph_1
\begin{pmatrix}
t_1\\ 
\vdots \\ 
t_p
\end{pmatrix}(1,t_1,\dots,t_p)\text{d}t_1\cdots \text{d}t_p\cdot\mu+o(1)\\
&\equiv \frac{3\alpha_{n}^2}{32\sqrt{\pi}h_2h_{1}^{p-2}}g'''(0\mid X_{0})f(X_{0})\mu^\top \Delta_2\mu+o(1)
\end{align*}
Hence,
$$\text{Var}(\mathbb{I}_1)=\frac{\alpha_{n}^2}{4\sqrt{\pi}nh_2^3h_{1}^{p}}g(0\mid X_{0})f(X_{0})\mu^\top \Delta_1\mu+o(1)=O\left(\frac{\alpha_{n}^2c^2}{nh_2^3h_{1}^{p}}\right).$$
As a relust,
\begin{align*}
    \mathbb{I}_1&=E(\mathbb{I}_1)+\text{Var}(\mathbb{I}_1)^{1/2}\\
    &=O\left(\alpha_n c\left(h_1^2+h_2^2\right)\right)+O\left(\frac{\alpha_{n}c}{\sqrt{nh_2^3h_{1}^{p}}}\right)=O(\alpha_{n}^2c)=O_p(\alpha_{n}^2c).
\end{align*}
Similar to the calculation process of mean and variance for $\mathbb{I}_1$, we have
\begin{align*}
E(\mathbb{I}_2)&=E\left[K_{h_1}\alpha_n^2(\mu^\top \bm{\text{X}}_{i}^{*})^2\left(-\frac{1}{2}g'''(0\mid\bm{X}_i)R(\bm{X}_i)+\frac{1}{2}g''(0\mid\bm{X}_i)+o_p(1)\right)\right]\\
&=-\frac{\alpha_n^2\mu h_1^2}{2}g'''(0\mid X_{0})f(X_{0})\int\cdots\int K(t_1,\dots,t_p)\begin{pmatrix}
1\\ 
t_1\\ 
\vdots \\ 
t_p
\end{pmatrix}\\
&\qquad \qquad \cdot (t_1,\dots,t_p)m^{(2)}(X_{0})\begin{pmatrix}
t_1\\ 
\vdots \\ 
t_p
\end{pmatrix}(1,t_1,\dots,t_p)\text{d}t_1\cdots \text{d}t_p\cdot\mu\\
&+\frac{\alpha_n^2\mu}{2}g''(0\mid X_{0})f(X_{0})\int\cdots\int K(t_1,\dots,t_p)\begin{pmatrix}
1\\ 
t_1\\ 
\vdots \\ 
t_p
\end{pmatrix}\\
&\qquad \qquad \cdot(1,t_1,\dots,t_p)\text{d}t_1\cdots \text{d}t_p\cdot\mu+o(1)\\
&\equiv \frac{\alpha_n^2}{2}g''(0\mid X_{0})f(X_{0})\mu^\top  \tilde{\Delta}_1\mu+o(1).
\end{align*}
Omits the variance of $\mathbb{I}_2$ calculation procedure. If $nh_1^ph_2^5\rightarrow \infty$, we have $$\mathbb{I}_2=\frac{\alpha_n^2}{2}g''(0\mid X_{0})f(X_{0})\mu \tilde{\Delta}_1\mu+o(1)$$ and 
$$\mathbb{I}_3=\frac{1}{n}\sum_{i=1}^{n}\left(-\frac{1}{6}\phi_{h_{2}}'''\left(\zeta _{i}\right)\alpha_{n}^{3}(\mu^\top \bm{\text{X}}_{i}^{*})^{3} \right)=o_p(\alpha_n^2).$$

Notice that and $\tilde{\Delta}_1$ is a positive-defined matrix, $\|\mu\|_2=c$ and $g''(0\mid X_{0})<0$, we can choose $c$ big enough such that the second term $\mathbb{I}_2$ dominates the others terms $\mathbb{I}_1$ and $\mathbb{I}_3$ with the probability $1-\eta$ for any given $\eta>0$. Then \eqref{promianeq} holds. Therefore, there exists a local maximizer $\hat{\theta}^*$ such that $\|\hat{\theta}^*-\theta^*\|_2\leq \alpha_n c$ with $\eta$ is smaller enough, where $\hat{\theta}^*=H\hat{\theta}=(\hat{b}_0, h_1\hat{b}_1,\dots, h_1\hat{b}_p).$ 

Based on \eqref{promianeq}, for any $\hat{\theta}^*$ such that $\|\hat{\theta}^*-\theta^*\|_2\leq \alpha_n c$ with $\eta$ is smaller enough, we have $\textup{L}(\hat{\theta}^*)<\textup{L}(\theta^*)$. In other words, when maximizing $\textup{L}(\cdot)$ in within a radius $\alpha_n c$ of $\theta^*$, the maximizer must be inside that neighboring region, that is $\|\hat{\theta}^*-\theta^*\|_2\leq \alpha_n c$ with $\eta$ is smaller enough. Therefore, 
$\|\hat{\theta}^*-\theta^*\|_2=O_{p}\left((nh_{1}^ph_{2}^{3})^{-1/2}+h_{1}^{2}+h_{2}^{2}\right)$ is completed.

Immediately, we can get
$$h_1\left \|\hat{\bm{b}}-\bm{b}\right\|_2=O_{p}\left((nh_{1}^ph_{2}^{3})^{-1/2}+h_{1}^{2}+h_{2}^{2}\right),$$
where $h_1\hat{\bm{b}}$ and $h_1\bm{b}$ are $2nd, \dots,(p+1)$ elements of $\hat{\theta}^*$ and $\theta^*$ respectively. 
\begin{align*}
    h_1\left \|\hat{\bm{b}}\hat{\bm{b}}^\top -\bm{b}\bm{b}^\top \right\|_F&= h_1\left \|\left(\hat{\bm{b}}-\bm{b}\right)\hat{\bm{b}}^\top +\bm{b}\left(\hat{\bm{b}}^\top \bm{b}^\top \right)\right\|_F\\
    &\leq h_1\left \|\left(\hat{\bm{b}}-\bm{b}\right)\hat{\bm{b}}^\top \right\|_F+h_1\left \|\bm{b}\left(\hat{\bm{b}}^\top \bm{b}^\top \right)\right\|_F\\
    &=h_1\left \|\hat{\bm{b}}-\bm{b}\right\|_2\left \|\hat{\bm{b}}^\top \right\|_2+h_1\left \|\bm{b}\right\|_2\left \|\hat{\bm{b}}-\bm{b}\right\|_2\\
    &=O_{p}\left((nh_{1}^ph_{2}^{3})^{-1/2}+h_{1}^{2}+h_{2}^{2}\right),
\end{align*}
Notice that $\hat{\bm{b}}$ and $\bm{b}$ rely on $\bm{X}_i$. According to above equation and condition (A5), we have 
$$h_1E \left(\left\|\hat{\bm{b}}\hat{\bm{b}}^\top -\bm{b}\bm{b}^\top \right\|_F\right)=O_{p}\left((nh_{1}^ph_{2}^{3})^{-1/2}+h_{1}^{2}+h_{2}^{2}\right).$$
By Jensen's  inequality,
$$\left\|E\left(\hat{\bm{b}}\hat{\bm{b}}^\top -\bm{b}\bm{b}^\top \right)\right\|_F\leq E \left(\left\|\hat{\bm{b}}\hat{\bm{b}}^\top -\bm{b}\bm{b}^\top \right\|_F\right).$$
Hence, 
$$h_1\left\|E\left(\hat{\bm{b}}\hat{\bm{b}}^\top -\bm{b}\bm{b}^\top \right)\right\|_F=O_{p}\left((nh_{1}^ph_{2}^{3})^{-1/2}+h_{1}^{2}+h_{2}^{2}\right).$$
Denote $E(\hat{\bm{b}}\hat{\bm{b}}^\top )$ is $\hat{\text{B}}$. By Lemma \ref{Zhidaolem}, immediately, 
$$\left\|\text{B}\text{B}^\top -\text{B}_0\text{B}_0^\top \right\|_F=0.$$
Then,
\begin{align*}
    h_1\left\|\hat{\text{B}}\hat{\text{B}}^\top -\text{B}_0\text{B}_0^\top \right\|_F&=h_1\left\|\hat{\text{B}}\hat{\text{B}}^\top -\text{B}\text{B}^\top +\text{B}\text{B}^\top -\text{B}_0\text{B}_0^\top \right\|_F\\
    &\leq h_1\left\|\hat{\text{B}}\hat{\text{B}}^\top -\text{B}\text{B}^\top \right\|_F\\
    &= h_1\left\|(\hat{\text{B}}-\text{B})\hat{\text{B}}^\top +\text{B}(\hat{\text{B}}-\text{B})^\top \right\|_F\\
    &\leq h_1\left\|\hat{\text{B}}-\text{B}\right\|_F\left\|\hat{\text{B}}^\top \right\|_2+h_1\left\|\text{B}\right\|_2\left\|(\hat{\text{B}}-\text{B})^\top \right\|_F\\
    &=O_{p}\left((nh_{1}^ph_{2}^{3})^{-1/2}+h_{1}^{2}+h_{2}^{2}\right),
\end{align*}
The proof is completed.
\end{proof}

\subsection*{Proof of Theorem 3}
\begin{proof}
By the rebuilding model \eqref{NewModel},
\eqref{TylEpa} and the definition of $R(\bm{X}_i)$, 
    let $$\hat{\gamma}_{i}=R(\bm{X}_i)-\left((\hat{b}_0-b_0)+\left(\hat{\bm{b}}-\bm{b}\right)^\top (\bm{X}_i-X_0)\right)=R(\bm{X}_i)-\left(\hat{\theta}^*-\theta^*\right)^\top \bm{\text{X}}_{i}^{*}.$$
    Then $$\textsl{y}_i-\hat{b}_0-\hat{\bm{b}}^\top (\bm{X}_i-X_0))=\varepsilon_i^*+\hat{\gamma}_{i}.$$
$\hat{\theta}^*$ is a solution satisfying following equation:
\begin{align*}
    \sum_{i=1}^{n}\bm{\text{X}}_{i}^{*}K_{h_1}(\bm{X}_i-\bm{X}_0)\phi_{h_2}'(\varepsilon_i^*+\hat{\gamma}_{i})=0.
\end{align*}
By the Taylor expansion, the above equation can be written as
\begin{equation}
\label{biaseq}
     \sum_{i=1}^{n}\bm{\text{X}}_{i}^{*}K_{h_1}(\bm{X}_i-\bm{X}_0)\left(\phi_{h_2}'(\varepsilon_i^*)+\phi_{h_2}''(\varepsilon_i^*)\hat{\gamma}_{i}+\frac{1}{2}\phi_{h_2}'''(\varepsilon_{i_{\xi}}^*)\hat{\gamma}_{i}^2\right)=0,
\end{equation}
where $\varepsilon_{i_{\xi}}^*$ is between $\varepsilon_i^*$ and $\varepsilon_i^*+\hat{\gamma}_{i}$. Next, Calculate the second term on the left side of the equation \eqref{biaseq}.
\begin{align*}
    &\sum_{i=1}^{n}\bm{\text{X}}_{i}^{*}K_{h_1}(\bm{X}_i-\bm{X}_0)\phi_{h_2}''(\varepsilon_i^*)\hat{\gamma}_{i}\\
&=\sum_{i=1}^{n}K_{h_1}\phi_{h_2}''(\varepsilon_i^*)R(\bm{X}_i)\bm{\text{X}}_{i}^{*}-\sum_{i=1}^{n}K_{h_1}\phi_{h_2}''(\varepsilon_i^*)\bm{\text{X}}_{i}^{*}\bm{\text{X}}_{i}^{*^\top }\left(\hat{\theta}^*-\theta^*\right)\equiv \mathbb{J}_1+\mathbb{J}_2.
\end{align*}
Similar to the proof of $\mathbb{I}_1$, $\mathbb{I}_2$ and $\mathbb{I}_3$, we have
\begin{align*}
    \mathbb{J}_1=\frac{n}{2}h_1^2g''(0\mid X_{0})f(X_0)\upsilon_{\textup{I}_2} +o_p(nh_1^2)
\end{align*}
and
\begin{align*}
    \mathbb{J}_2=-ng''(0\mid X_{0})f(X_0)\tilde{\Delta}_1\left(\hat{\theta}^*-\theta^*\right)+o_p(n\|\hat{\theta}^*-\theta^*\|_2).
\end{align*}
Because $\|\hat{\theta}^*-\theta^*\|_2=O_p\left((nh_{1}^ph_{2}^{3})^{-1/2}+h_{1}^{2}+h_{2}^{2}\right)=O_p(\alpha_n)$, 
\begin{align*}
    \mathop{\text{Sup}}_{i:\frac{\|\bm{X}_i-X_0\|_2}{h_1}\leq 1}|\hat{\gamma}_i|&\leq \mathop{\text{Sup}}_{i:\frac{\|\bm{X}_i-X_0\|_2}{h_1}\leq 1} | R(\bm{X}_i)|+|(\hat{\theta}^*-\theta^*)^\top \bm{\text{X}}_{i}^{*}|\\
    &=O_p(h_1^2+\|\hat{\theta}^*-\theta^*\|_2)=O_p(\|\hat{\theta}^*-\theta^*\|_2)=O_p(\alpha_n)=o_p(1).
\end{align*}
Hence,
\begin{align}
\label{supgamma2}
     \mathop{\text{Sup}}_{i:\frac{\|\bm{X}_i-X_0\|_2}{h_1}\leq 1}|\hat{\gamma}_i^2|=o_p(1)O_p(\|\hat{\theta}^*-\theta^*\|_2)=o_p(\|\hat{\theta}^*-\theta^*\|_2).
\end{align}
For the third term on the left side of the equation \eqref{biaseq}, by $\mathbb{I}_3$, \eqref{supgamma2} and $\|\hat{\theta}^*-\theta^*\|_2=O_p(\alpha_n)$,  we can obtain
\begin{align*}
    \frac{1}{2}\sum_{i=1}^{n}\bm{\text{X}}_{i}^{*}K_{h_1}\phi_{h_2}'''(\varepsilon_{i_{\xi}}^*)\hat{\gamma}_{i}^2=O_p(\alpha_n^2)\left(\frac{1}{2}\sum_{i=1}^{n}\bm{\text{X}}_{i}^{*}K_{h_1}\phi_{h_2}'''(\varepsilon_{i_{\xi}}^*)\right)
=o_p(n\alpha_n)=o_p(\mathbb{J}_2).
\end{align*}
Let $V_n= \sum_{i=1}^{n}\bm{\text{X}}_{i}^{*}K_{h_1}(\bm{X}_i-\bm{X}_0)\phi_{h_2}'(\varepsilon_i^*)$, then
\begin{align}
\label{hattheta-theta}
    \hat{\theta}^*-\theta^*=\frac{\tilde{\Delta}_1^{-1}V_n}{ng''(0\mid X_{0})f(X_0)}+\frac{h_1^2\tilde{\Delta}_1^{-1}\upsilon_{\textup{I}_2}}{2}+o_p(1).
\end{align}
According to above equation, the next step is to calculate the mean and variance of $V_n/n$. Following  the above calculations, we have
\begin{align*}
    E\left(\frac{1}{n}V_n\right)=-\frac{h_2^2}{2}g'''(0\mid X_0)f(X_0)\upsilon_{\textup{I}_1}+o_p(1),
\end{align*}
and
\begin{align*}
\textup{Var}\left(\frac{1}{n}V_n\right)=\frac{1}{4nh_1^ph_2^3}g(0\mid X_0)f(X_0)\Delta_1+o_p(1).
\end{align*}
By the central limit theorem, we have
\begin{align}
\label{Vnasynorm}
    \sqrt{\frac{2h_1^ph_2^3}{n}}\left(V_n+\frac{h_2^2}{2}g'''(0\mid X_0)f(X_0)\upsilon_{\textup{I}_1}\right)\stackrel{L}{\longrightarrow} \mathrm{N}(0, g(0\mid X_0)f(X_0)\Delta_1).
\end{align}
Furthermore, the asymptotic bias $\tilde{\mathbf{b}}$ and variance of $\hat{\theta}^*$ can be given by 
\begin{align}
\label{bias}
    \tilde{\mathbf{b}}\approx \frac{h_1^2\tilde{\Delta}_1^{-1}\upsilon_{\textup{I}_2}}{2}-\frac{h_2^2g'''(0\mid X_0)\tilde{\Delta}_1^{-1}\upsilon_{\textup{I}_1}}{2g''(0\mid X_{0})}
\end{align}
and
\begin{align}
\label{asyVartheta}
    \textup{Var}(\hat{\theta}^*)\approx \frac{g(0\mid X_0)\tilde{\Delta}_1^{-1}\Delta_1\tilde{\Delta}_1^{-1}}{4nh_1^ph_2^3g^{''^{2}}(0\mid X_0)f(X_0)}.
\end{align}
Using Slutsky's theorem, it follows from \eqref{hattheta-theta}, \eqref{Vnasynorm}, \eqref{bias} and \eqref{asyVartheta} that
\begin{align*}
    \left[\textup{Var}\left(\hat{\theta}^*\right)\right]^{-\frac{1}{2}}\left(\hat{\theta}^*+\frac{h_2^2g'''(0\mid X_0)\tilde{\Delta}_1^{-1}\upsilon_{\textup{I}_1}}{2g''(0\mid X_{0})}-\tilde{\mathbf{b}} \right)\stackrel{L}{\longrightarrow} \mathrm{N}(0, I_{p+1}).
\end{align*}

\end{proof}

\bibliographystyle{apalike}
\bibliography{refs.bib}
\end{document}